\documentclass[journal]{IEEEtran}

\usepackage{stfloats}
\usepackage{algorithm}  
\usepackage{algpseudocode}  
\usepackage{amsmath}

\usepackage{graphicx}
\usepackage[]{algpseudocode}
\usepackage{algorithmicx,algorithm}
\usepackage{amsmath}
\usepackage{titlesec}
\usepackage{multirow}  
\usepackage{amsthm,amsmath,amssymb}
\usepackage{graphicx}
\usepackage{float}
\usepackage{subfigure}
\usepackage{mathrsfs}
\usepackage{url}
\usepackage{array}
\newcolumntype{C}[1]{>{\centering}p{#1}}
\usepackage{makecell}
\usepackage{diagbox}
\usepackage{amsmath}

\usepackage{gensymb}
\usepackage{cite}
\usepackage{color}

\ifCLASSOPTIONcompsoc
 \usepackage[caption=false,font=normalsize,labelfont=sf,textfont=sf]{subfig}
\else
 \usepackage[caption=false,font=footnotesize]{subfig}
\fi

\UseRawInputEncoding
\begin{document}
\theoremstyle{definition}
\newtheorem{remark}{Remark}
\newtheorem{theorem}{Theorem} 
\newtheorem{definition}[theorem]{Definition} 
\newtheorem{lemma}{Lemma} 
\newtheorem{corollary}[theorem]{Corollary}
\newtheorem{example}{Example} 
\newtheorem{proposition}[theorem]{Proposition}

\vspace{-15mm}
\title{Electromagnetic Property Sensing Based on Diffusion Model in ISAC System
}


\author{   
Yuhua Jiang, Feifei Gao, Shi Jin, and Tie Jun Cui

\thanks{
Y. Jiang and F. Gao are with Institute for Artificial Intelligence, Tsinghua University (THUAI), 
State Key Lab of Intelligent Technologies and Systems, Tsinghua University, 
Beijing National Research Center for Information Science and Technology (BNRist), Beijing, P.R. China (email: jiangyh221600@gmail.com, feifeigao@ieee.org).


S. Jin is with the National Mobile Communications Research 
Laboratory, Southeast University, Nanjing 210096, China (e-mail: jinshi@seu.edu.cn).

T. J. Cui is with the State Key Laboratory of Millimeter Waves, Southeast University, Nanjing 210096, China (e-mail: tjcui@seu.edu.cn).



}
}

\maketitle
\vspace{-15mm}
\begin{abstract}
Integrated sensing and communications (ISAC) has opened up numerous game-changing opportunities for future wireless systems. 
In this paper, we develop a novel ISAC scheme that utilizes the diffusion model to sense the electromagnetic (EM) property of the target in a predetermined sensing area. 
Specifically, we first estimate the sensing channel by using both the communications and the sensing signals echoed back from the target. 
Then we employ the diffusion model to generate the point cloud that represents the target and thus enables 3D visualization of the target's EM property distribution. 
In order to minimize the mean Chamfer distance (MCD) between the ground truth and the estimated point clouds, we further design the communications and sensing beamforming matrices under the constraint of a maximum transmit power and a minimum communications achievable rate for each user equipment (UE). 
Simulation results demonstrate the efficacy of the proposed method in achieving high-quality reconstruction of the target's shape, relative permittivity, and conductivity. 
Besides, the proposed method can sense the EM property of the target effectively in any position of the sensing area.


\end{abstract}


\IEEEpeerreviewmaketitle


\section{Introduction} 

Recently, the concept of integrated sensing and communications (ISAC) has attracted significant interest from academy and industry professionals, particularly for its potential in the sixth-generation (6G) wireless networks \cite{isac9,overview_ourgroup}.
Compared with the traditional frequency-division sensing and communications (FDSAC) approach that requires separate frequency bands and infrastructure for each function, 
ISAC could concurrently share time, frequency, power, and hardware resources for both communications and sensing functions, 
and is anticipated to outperform FDSAC in terms of spectrum utilization, energy efficiency, and hardware requirements \cite{isac1,mypaper3,isac2}. 
Additionally, ISAC can be integrated with other emerging technologies, such as reconfigurable intelligent surfaces (RIS), to enhance the performance of sensing and communications systems \cite{ mypaper}.



With its numerous advantages, ISAC is expected to be crucial in a variety of emerging applications such as vehicle to everything (V2X) \cite{v2x}, extended reality (XR) \cite{xr}, smart homes \cite{iot}, and digital twins that seamlessly connects the tangible world with its virtual counterpart in the realm of communications \cite{twin}.
In contrast to the image-based digital twins that focus on shape and spatial positioning, the digital twins for communications systems are tasked with the intricate reconstruction of communications pathways and the management of channel-specific challenges. 
In addition to the location, velocity, and shape, the precise values of the electromagnetic (EM) property of real-world entities are also significant.  
These attributes are pivotal as they dictate the physical phenomena such as diffraction, reflection, and scattering of EM waves. 
Moreover, sensing the EM property is essential in human body examination, which meets the escalating needs in societal security and the medical field.

Although ISAC has demonstrated important achievements in localization \cite{isac8}, tracking \cite{isac3}, imaging \cite{RIS_image}, and various other applications \cite{isac1, mypaper3, isac2, mypaper}, EM property sensing within ISAC framework has been barely studied. 
To the best of the authors' knowledge, only two works \cite{mypaper4,mypaper5} implement EM property sensing through mesh-based methods. 
However, the methods in \cite{mypaper4} and \cite{mypaper5} only address 2D scenarios and are not suitable for 3D scenarios,
because the computational cost and memory requirements in 3D scenarios escalate significantly. 
Additionally, the complexity of mesh generation and the need for more sophisticated meshing algorithms further complicate the extension of \cite{mypaper4} and \cite{mypaper5} to 3D scenarios.

Hence, reducing the complexity of ISAC algorithms in 3D scenarios is crucial to accurately reflecting the propagation of EM waves in the real world.
\textcolor{black}{
Recent studies have applied the artificial intelligence (AI) algorithms in ISAC systems to address the high-computational-complexity problem.}    
For instance, the work by Zhu et al. \cite{isasc_ai1} proposed an AI-enabled STAR-RIS aided ISAC system for secure communications.
Additionally, Zhu et al. \cite{isasc_ai2} provided an overview of how AI can be pushed to the wireless network edge, highlighting the opportunities for integrated sensing, communications, and computation towards 6G.
Moreover, Wen et al. \cite{isasc_ai3} investigated task-oriented sensing, computation, and communications integration for multi-device edge AI.
Wang et al. \cite{isasc_ai4} explored the generative AI (GAI) in ISAC systems from the physical layer perspective.

\textcolor{black}{
However, transferring conventional AI methods to 3D EM property sensing presents specific challenges due to the ill-posed nature of the inverse problems involved in EM property sensing. Traditional AI methods may face significant hurdles when tasked with these complex inverse problems \cite{li2018deepnis}.
One of the primary difficulties is the high degree of non-linearity and the presence of multiple scattering effects in EM wave propagation. These factors contribute to the ill-posedness of the problem, making it difficult to accurately infer the underlying EM property from the observed data. Traditional AI models exhibit high computational complexity, poor solution effectiveness, and significant sensitivity to noise when addressing complex EM inverse scattering problems \cite{zhang2022probabilistic,guo2023physics}. 
Moreover, small errors in the input data can lead to large deviations in the estimated EM property, which is particularly problematic for traditional AI methods.
This sensitivity to noise and measurement errors can yield unreliable or inconsistent results, undermining the effectiveness of EM property sensing in practical applications.}%

\textcolor{black}{%
One promising AI approach to overcome these challenges is the diffusion model.  
Recently, GAI has become increasingly popular due to its versatility and capability to perform a wide range of tasks.
Among the GAI models, the diffusion model stands out as a particularly effective approach.
Unlike other GAI methods, the diffusion model offers enhanced flexibility and efficiency, making it well-suited for EM property sensing 
\cite{tutorial_on_Diffusion, VLB_diffusion}.
The diffusion model's robustness to noise ensures accurate data collection,
while its adaptability allows the model to be fine-tuned for various applications. The diffusion model's efficiency is crucial for real-time processing, and its scalability handles large datasets effectively. Moreover, diffusion models can integrate prior knowledge and capture complex relationships, which provides a comprehensive understanding of EM property. Their continuous learning capability and compatibility with other technologies make them versatile tools in the field of EM property sensing.
}

In this paper, we develop a novel ISAC scheme that leverages the diffusion model to accurately sense the EM property of a target within a predetermined sensing area. We first utilize both communications and sensing signals that are reflected back from the target to estimate the sensing channel. 
Once the sensing channel is estimated, we employ the diffusion model to generate a detailed point cloud representation of the target. This point cloud provides a clear 3D visualization of the EM property distribution, including critical parameters such as shape, relative permittivity, and conductivity. To ensure high fidelity in these visualizations, we design the communications and sensing beamforming matrices to minimize the mean Chamfer distance between the ground truth and the estimated point clouds. This optimization is performed under the constraints of a maximum transmit power and a minimum achievable communications rate for each user equipment (UE).
Simulation results validate the efficacy of the proposed method and demonstrate its capability to achieve high-quality reconstruction of the target's EM property. Additionally, the  proposed method proves to be effective in sensing the EM property of the target regardless of its position within the sensing area. This robustness highlights the potential of the  proposed ISAC scheme to be applied in various practical scenarios where precise EM property sensing is required. 


The rest of this paper is organized as follows. 
Section~\uppercase\expandafter{\romannumeral2} presents the ISAC system model.
Section~\uppercase\expandafter{\romannumeral3} elaborates the formulation of EM scattering. 
Section~\uppercase\expandafter{\romannumeral4} describes the diffusion probabilistic model for EM property sensing.
Section~\uppercase\expandafter{\romannumeral5} proposes the approach to designing the beamforming matrices with tradeoff between sensing and communications.
Section~\uppercase\expandafter{\romannumeral6} provides the numerical simulation results, and Section~\uppercase\expandafter{\romannumeral7} draws the conclusion.

Notations: Boldface denotes a vector or a matrix; $j$ corresponds to the imaginary unit; $(\cdot)^H$, $(\cdot)^\top$, and $(\cdot)^*$ represent Hermitian, transpose, and conjugate, respectively; 
$\odot$ denotes the Hadamard product; 
$\otimes$ denotes the Kronecker product; 
$\mathrm{vec}(\cdot)$ and $\mathrm{unvec}(\cdot)$ denote the vectorization and unvectorization operation; 
$\nabla$ denotes the nabla operator; 
$\textrm{diag}(\mathbf{a})$ denotes the diagonal matrix whose diagonal elements are the elements of $\mathbf{a}$; 
$\mathbf{I}$ denote the identity matrix with compatible dimensions; 
$\left\Vert\mathbf{a}\right\Vert_2$ denotes $\ell2$-norm of the vector $\mathbf{a}$; 
$\left\Vert\mathbf{A}\right\Vert_F$ denotes Frobenius-norm of the matrix $\mathbf{A}$; 
$\text{tr}(\mathbf{A})$ denotes the trace of the matrix $\mathbf{A}$; 
$\left| \cdot \right|$ denotes the 
element-wise absolute value of complex vectors or matrices; 
\color{blue}
$\mathbb{E}_{q\left(\mathbf{a}\right)} f (\mathbf{a}) \triangleq \int f(\mathbf{a}) q(\mathbf{a}) \, \mathrm{d} \mathbf{a} $ denotes the expected value of $f (\mathbf{a})$ with respect to the probability distribution $q\left(\mathbf{a}\right)$;
\color{black}
$\mathbf{A} \succeq \mathbf{0}$ indicates that the matrix $\mathbf{A}$ is positive semi-definite; 
the distribution of a real-valued Gaussian random vector with mean $\boldsymbol{\mu}$ and 
covariance matrix $\mathbf{A}$ is denoted as $\mathcal{N} (\boldsymbol{\mu}, \mathbf{A})$;
\color{blue}
the distribution of a real-valued random vector $\mathbf{a}$ subject to $\mathcal{N} (\boldsymbol{\mu}, \mathbf{A})$ is denoted as $\mathcal{N} (\mathbf{a} \mid \boldsymbol{\mu}, \mathbf{A})$;
\color{black}
the distribution of a circularly symmetric complex Gaussian (CSCG) random vector with mean $\boldsymbol{\mu}$ and 
covariance matrix $\mathbf{A}$ is denoted as $\mathcal{C N} (\boldsymbol{\mu}, \mathbf{A})$. 






 


\begin{figure}[t]
  \centering  \centerline{\includegraphics[width = 5.9cm]{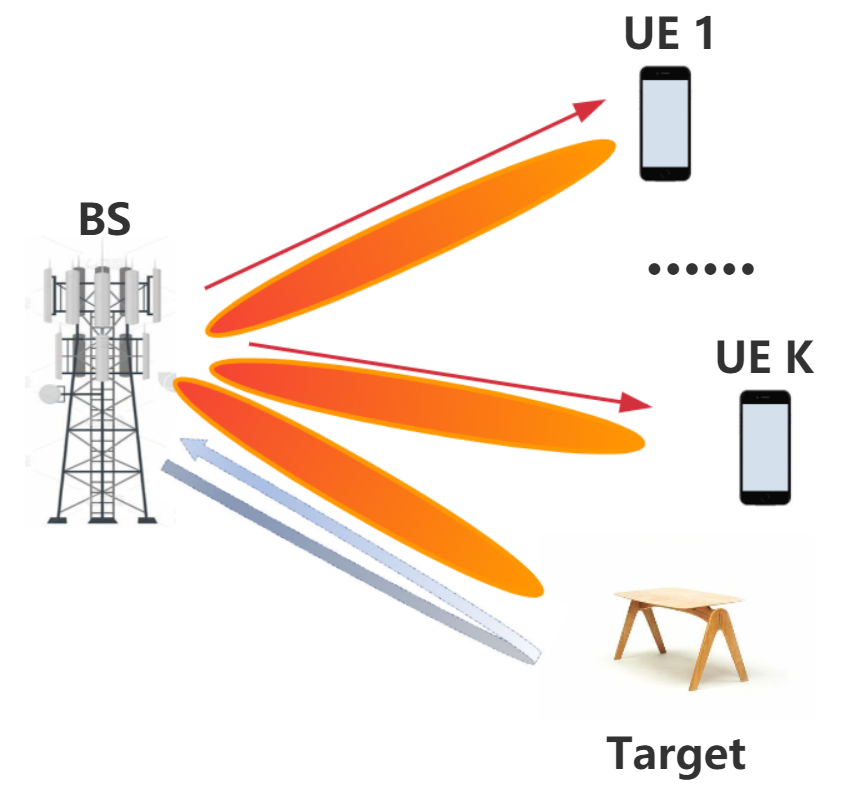}}
  \caption{The system model for simultaneous communications information transmission and EM property sensing. 
  } 
  \label{system_model}
\end{figure}

\section{System Model}
As illustrated in Fig. 1, consider a multi-antenna ISAC system for simultaneous information transmission and the target's EM property sensing. 
The system includes a base station (BS) equipped with $N_t$ transmitting antennas and $N_r$ receiving antennas, $K$ UEs each equipped with a single antenna, as well as a target to be sensed. 
The BS transmits communications signals to the $K$ UEs simultaneously and utilizes the received echo signals to sense the EM property of the target. 
The scenario is also known as the mono-static ISAC system.

The BS adopts fully digital precoding structure where the number of RF chains $N_{RF}$ is equal to the number of antennas $N_t$.
The frequency of the signals is denoted by $f$, and the length of the transmitted symbols is denoted by $L$.
We assume a quasi-static environment where the communications and sensing channels remain unchanged throughout the ISAC period. 


Suppose the $l$-th transmit signal at the BS is expressed as
\begin{align}
\mathbf{x}(l) &  = \mathbf{W}_{\mathrm{s}} \mathbf{s}(l)+\mathbf{W}_{\mathrm{c}} \mathbf{c}(l) \nonumber \\
& =\left[\begin{array}{ll}
\mathbf{W}_{\mathrm{s}},  \mathbf{W}_{\mathrm{c}}
\end{array}\right]\left[\begin{array}{ll}
\mathbf{s}(l)^{\top},  \mathbf{c}(l)^{\top}
\end{array}\right]^{\top}=\mathbf{W} \hat{\mathbf{x}}(l),
\end{align} 
where $\mathbf{s}(l) \in \mathbb{C}^{N_t \times 1}$ denotes the dedicated sensing signal and $\mathbf{c}(l) \in \mathbb{C}^{K \times 1}$ denotes the transmission symbols intended to $K$ UEs. 
Moreover, $\mathbf{W}_{\mathrm{s}}=\left[\mathbf{w}_{\mathrm{s}, 1}, \mathbf{w}_{\mathrm{s}, 2}, \cdots, \mathbf{w}_{\mathrm{s}, N_t}\right] \in \mathbb{C}^{N_t \times N_t}$ and $\mathbf{W}_{\mathrm{c}}=\left[\mathbf{w}_{\mathrm{c}, 1}, \mathbf{w}_{\mathrm{c}, 2}, \cdots, \mathbf{w}_{\mathrm{c}, K}\right] \in \mathbb{C}^{N_t \times K}$ represent the digital beamforming matrices for sensing and communications, respectively.
Besides, $\mathbf{W} = \left[
\mathbf{W}_{\mathrm{s}}, \mathbf{W}_{\mathrm{c}} \right]$ denotes the equivalent transmit beamforming matrix.
We assume that the sensing signal is generated pseudo-randomly, which satisfies $\mathbb{E}[\mathbf{s}(l)]=\mathbf{0}$ and $\mathbb{E}\left[\mathbf{s} (l) \mathbf{s}^{H}(l)\right] = \mathbf{I}_{N_t} $. 
We also assume that the transmitted communications symbol $\mathbf{c}(l)$ satisfies $\mathbf{c}(l) \sim \mathcal{C} \mathcal{N}\left(\mathbf{0}, \mathbf{I}_K \right) $, while the sensing and communications signals are mutually uncorrelated. 
Thus, the transmit signal covariance matrix is given by
\begin{align}
\mathbf{S}_x = \mathbb{E}\left[\mathbf{x}(l) \mathbf{x}^{H}(l)\right]=\mathbf{W} \mathbf{W}^{H}= \mathbf{R}_s + \sum_{k=1}^K \mathbf{R}_k ,
\label{RS}
\end{align}
where the Gram matrix for the dedicated sensing is defined as $ \mathbf{R}_s \triangleq \mathbf{W}_s \mathbf{W}_{s}^{H}$,
and the rank-one Gram matrix for the $k$-th UE's communications is defined as $\mathbf{R}_k \triangleq \mathbf{w}_{c, k} \mathbf{w}_{c, k}^{H}$. 

Assume that the BS operates under a maximum  transmit power $P$, i.e.,  $\operatorname{tr}\left(\mathbf{S}_x\right) \leq P$. Let $\mathbf{X}=[\mathbf{x}(1), \cdots, \mathbf{x}(L) ]$ denote the transmitted signal over $L$ symbols. 
By assuming that $L$ is sufficiently large, the sample covariance matrix of $\mathbf{X}$ can be approximated as the statistical covariance matrix $\mathbf{S}_x$, i.e.,
$ \frac{1}{L} \mathbf{X} \mathbf{X}^H \approx \mathbf{S}_x$.

\subsection{Communications Model}
Assume each UE is equipped with a single antenna. 
Let $\mathbf{h}_k \in \mathbb{C}^{N_t \times 1}$ denote the channel vector from the BS to the $k$-th UE. The received signal at the $k$-th UE without sensing interference cancellation is given by 
\begin{align}
y_k(l)=\mathbf{h}_k^H \mathbf{x}(l)+z_k(l), \quad l = 1, \cdots, L , 
\label{receive}
\end{align}
where $z_k(l) \sim \mathcal{C N}\left(0, \sigma_k^2\right)$ denotes the noise at the $k$-th UE. 
Based on the received signal in (\ref{receive}), the received signal-to-interference-plus-noise ratio (SINR) at the $k$-th UE is 
\begin{align}
\gamma_k = \frac{\mathbf{h}_k^H \mathbf{R}_k \mathbf{h}_k }{\sigma_k^2 + \mathbf{h}_k^H (\mathbf{R}_s + \sum_{k' \neq k} \mathbf{R}_{k'}) 
\mathbf{h}_k} .
\label{gamma}
\end{align}

Substituting (\ref{RS}) into (\ref{gamma}), the achievable rate of the communications channel for the $k$-th UE with multi-UE and sensing interference is given by 
\begin{align}
R_k = \log _2\left(1+\frac{\mathbf{h}_k^H \mathbf{R}_k \mathbf{h}_k }{\sigma_k^2 + \mathbf{h}_k^H (\mathbf{S}_x - \mathbf{R}_k) 
\mathbf{h}_k} \right) .
\end{align}  

\subsection{Sensing Model}
Assume the BS adopts its $N_r$ receiving antennas to obtain the echo signals and then estimates the EM property of the target.
In this case, the received target echo signals at the BS is expressed as 
\begin{equation}
\mathbf{Y}=\mathbf{H}_s \mathbf{X}+\mathbf{Z},
\label{sensing_channel}
\end{equation}
where $\mathbf{Y}=[\mathbf{y}(1), \cdots, \mathbf{y}(L)]$, $\mathbf{Z}=[\mathbf{z}(1), \cdots, \mathbf{z}(L)]$ denotes the noise at the BS receiver with $\mathbf{z}(l) \sim \mathcal{C N}(\mathbf{0}, \sigma_s^2\mathbf{I}_{N_r})$, 
and $\mathbf{H}_s$ denotes the sensing channel matrix.
Let us vectorize (\ref{sensing_channel}) as
\begin{equation}
\tilde{\mathbf{y}}=\left(\mathbf{X}^\top \otimes \mathbf{I}_{N_r}\right) \tilde{\mathbf{h}}_s+\tilde{\mathbf{z}},
\label{vec1}%
\end{equation}
where $\tilde{\mathbf{y}}=\operatorname{vec}(\mathbf{Y}) \in \mathbb{C}^{N_r L \times 1}, \tilde{\mathbf{h}}_s=\operatorname{vec}(\mathbf{H}_s) \in \mathbb{C}^{N_r N_t \times 1}$, and $\tilde{\mathbf{z}}=\operatorname{vec}(\mathbf{Z}) \in \mathbb{C}^{N_r L \times 1}$. It follows from (\ref{vec1}) that  $\tilde{\mathbf{y}} \sim \mathcal{C N}\left(\left(\mathbf{X}^\top \otimes \mathbf{I}_{N_r}\right)  \tilde{\mathbf{h}}_s, \sigma_s^2 \mathbf{I}_{L N_r}\right)$.

In order to extract the EM property of the target, the BS first needs to estimate $N_r N_t$ complex parameters in $\tilde{\mathbf{h}}_s$ or $\mathbf{H}_s$. 
It has been proven in \cite{crb} that the Cram\'er-Rao bound (CRB) matrix of $\tilde{\mathbf{h}}_s$ is 
\begin{align}
\textrm{CRB}&=\left[\left(\mathbf{X}^\top \otimes \mathbf{I}_{N_r}\right)^H\left(\sigma_s^2 \mathbf{I}_{L N_r}\right)^{-1}\left(\mathbf{X}^\top \otimes \mathbf{I}_{N_r}\right)\right]^{-1}\nonumber\\
&=\sigma_s^2\left(\mathbf{X}^* \mathbf{X}^\top \otimes \mathbf{I}_{N_r}\right)^{-1} \stackrel{(\mathrm{a})}{=} \frac{\sigma_s^2}{L}\left(\boldsymbol{S}_x^\top \otimes \mathbf{I}_{N_r}\right)^{-1},
\label{crb1}
\end{align}
where $\stackrel{(\mathrm{a})}{=}$ comes from the approximation $\frac{1}{L} \mathbf{X} \mathbf{X}^H \approx \mathbf{S}_x$.
We then adopt the trace of CRB as the scalar criterion to estimate $\tilde{\mathbf{h}}_s$, i.e., 
\begin{align}
\operatorname{tr}(\textrm{CRB})=\frac{N_r \sigma_s^2}{L} \operatorname{tr}\left(\boldsymbol{S}_x^{-1}\right) .
\label{crb2}
\end{align}

The least square (LS) method is applied as
\begin{align}
\hat{\tilde{\mathbf{h}}}_{s}=\left(\mathbf{X}^* \mathbf{X}^\top \otimes \mathbf{I}_{N_r}\right)^{-1}\left(\mathbf{X}^\top \otimes \mathbf{I}_{N_r}\right)^H \tilde{\mathbf{y}} ,
\label{crb3}
\end{align}
and there is $\hat{\tilde{\mathbf{h}}}_{s} \sim \mathcal{C N}\left(\tilde{\mathbf{h}}_{s}, \sigma_s^2\left(\mathbf{X}^*  \mathbf{X}^\top \otimes \mathbf{I}_{N_r}\right)^{-1}\right)$.
Consequently, $\hat{\mathbf{H}}_s = \operatorname{unvec}(\hat{\tilde{\mathbf{h}}}_{s})$ is the minimum variance unbiased estimated sensing channel that can achieve the CRB in (\ref{crb1}). 

\section{EM Scattering Formulation}
Assume the EM property of the target is isotropic. 
Sensing the EM property involves reconstructing the contrast function $\chi(\mathbf{r})$ defined as 
the difference in complex relative permittivity between the target and the background medium (the air here). Given that the relative permittivity and conductivity of air are approximately equal to $1$ and $0$ Siemens/meter (S/m), we can formulate the contrast function \cite{operator,born2} as
\begin{align}
\chi(\mathbf{r}) = \epsilon_r(\mathbf{r}) - \frac{j \sigma(\mathbf{r})}{\epsilon_0 \omega} - 1,
\label{contrast} 
\end{align}
where $\epsilon_r(\mathbf{r})$ denotes the real relative permittivity at point $\mathbf{r}$, $\sigma(\mathbf{r})$ denotes the conductivity at point $\mathbf{r}$, $\omega=2 \pi f$ denotes the angular frequency of the EM waves, and  $\epsilon_0$ is the vacuum permittivity. 
We then aim to reconstruct the distributions of both  $\epsilon_r(\mathbf{r})$ and $\sigma(\mathbf{r})$. 

Assume the electric fields have an $ e^{-j \omega t}$ time dependence throughout the paper.
Let $\lambda = c/f$ denote the wavelength and $k_b = 2\pi/ \lambda$ denote the wave number in the background medium. 
\color{blue}
The physical quantity at position $\mathbf {r}$ when the $l$-th signal is transmitted is represented by $(\mathbf {r} , l )$. 
Let $\mathbf{E}^t (\mathbf {r} , l ) \in \mathbb{C}^{3\times 1}$ and  
$\mathbf{E}^i ({\mathbf {r, l}}) \in \mathbb{C}^{3\times 1}$ denote the total electric field and the incident electric field
in $x$, $y$, and $z$ axis, respectively. 
Since the incident electric field is linearly induced by the currents on the transmitting antennas, there is a matrix $\mathbf{A} ({\mathbf {r}}) \in \mathbb{C}^{3 \times N_t}$ that linearly maps $\mathbf{x}(l)$ to $\mathbf{E}^i ({\mathbf {r} , l })$, i.e.,
\begin{align}
\mathbf{E}^i ({\mathbf {r}, l}) = \mathbf{A} ({\mathbf {r}}) \mathbf{x}(l) . 
\label{h1}
\end{align}
\color{black}
When the target is illuminated by an incident field $\mathbf{E}^i ({\mathbf {r}, l})$,
$\mathbf{E}^i ({\mathbf {r}, l})$ and $\mathbf{E}^t ({\mathbf {r}, l})$ will satisfy the homogeneous and inhomogeneous vectorial wave equations, respectively \cite{emscat_polarized}: 
\begin{align}
\nabla \times \nabla \times \mathbf{E}^i ({\mathbf {r}, l}) - k_b^2
\mathbf{E}^i ({\mathbf {r}, l}) &= \mathbf{0} , \label{inc} \\
\nabla \times \nabla \times \mathbf{E}^t ({\mathbf {r}, l}) - k_b^2 
\mathbf{E}^t ({\mathbf {r}, l}) &= k_b^2 \chi(\mathbf{r}) \mathbf{E}^t ({\mathbf {r}, l}) . \label{tot}
\end{align}

Suppose the target is located within the region $D$. 
In order to solve (\ref{inc}) and (\ref{tot}), 
the total electric field within $D$ can be formulated by the 3D Lippmann-Schwinger equation as \cite{emscat_polarized, lipp, lipp2}  
\begin{equation}
\mathbf{E}^t\left(\mathbf{r}, l \right) = \mathbf{E}^i\left(\mathbf{r}, l \right) + k_b ^2 \iiint_D \overline{\overline{\mathbf{G}}}\left(\mathbf{r}, \mathbf{r}^{\prime}\right) \chi\left(\mathbf{r}^{\prime}\right) \mathbf{E}^t \left(\mathbf{r}^{\prime}, l\right) \mathrm{d} \mathbf{r}^{\prime},
\label{lipp0}%
\end{equation}
where $\overline{\overline{\mathbf{G}}}\left(\mathbf{r}, \mathbf{r}^{\prime}\right) \in \mathbb{C}^{3\times 3}$ is the dyadic electric field Green's function that satisfies
\begin{equation}
\nabla \times \nabla \times \overline{\overline{\mathbf{G}}}\left(\mathbf{r}, \mathbf{r}^{\prime}\right) - k_b^2 \overline{\overline{\mathbf{G}}}\left(\mathbf{r}, \mathbf{r}^{\prime}\right) = 
\mathbf{I}_3 \delta\left(\mathbf{r}-\mathbf{r}^{\prime}\right) .
\end{equation}
Meanwhile, $\overline{\overline{\mathbf{G}}}\left(\mathbf{r}, \mathbf{r}^{\prime}\right)$ can be formulated as \cite{green}
\begin{align}
&\overline{\overline{\mathbf{G}}}\left(\mathbf{r},\mathbf{r}^{\prime}\right)=\left(\mathbf{I}_3+\frac{\nabla \nabla}{k_b^2}\right) g\left(\mathbf{r}, \mathbf{r}^{\prime}\right)\nonumber\\
&=\!\!\left[\!\!\left(\frac{3}{k_b^2 R'^2} - \frac{3 j}{k_b R'}-1\!\!\right) \! \hat{\mathbf{r}} \hat{\mathbf{r}}^\top\!\!\!\!-\!\!\left(\frac{1}{k_b^2 R'^2} - \frac{j}{k_b R'}-1\!\!\right)\!\! \mathbf{I}_3\right] \!\!g\! \left(\mathbf{r}, \mathbf{r}^{\prime}\right). \label{RR}%
\end{align}
In (\ref{RR}), $R'$ is the distance defined as $R' \triangleq \|\mathbf{r} - \mathbf{r}^{\prime} \|_2$, 
$\hat{\mathbf{r}}\in \mathbb{R}^{3 \times 1}$ is defined as the unit vector from $\mathbf{r}^{\prime}$ to $\mathbf{r}$, and $g\left(\mathbf{r}, \mathbf{r}^{\prime}\right)$ is the scalar Green's function defined as $g\left(\mathbf{r}, \mathbf{r}^{\prime}\right) \triangleq \frac{\exp(j k_b R')}{4 \pi R'}$ \cite{green}. 

The echo electric field at the BS's receiver scattered back from the target can then be formulated as  \cite{lipp, lipp2}
\begin{equation}
\mathbf{E}^s\left(\mathbf{r}_n, l \right) = k_b ^2 \iiint_D \overline{\overline{\mathbf{G}}}\left(\mathbf{r}_n, \mathbf{r}^{\prime}\right) \chi\left(\mathbf{r}^{\prime}\right) \mathbf{E}^t \left(\mathbf{r}^{\prime}, l\right) \mathrm{d} \mathbf{r}^{\prime},  
\label{lipp1}%
\end{equation}
where $\mathbf{r}_n$ denotes the position of the $n$-th receiving antenna. 
Suppose the receiver can only measure the scalar electric field component
in the direction represented by the unit vector $\hat{\mathbf{p}}\in \mathbb{R}^{3\times 1}$. 
\color{blue}
The $l$-th received echo signals can be formulated as 
\begin{align}
\mathbf{y}(l) = G_r \left[\mathbf{E}^s\left(\mathbf{r}_1 , l \right) , \cdots, \mathbf{E}^s\left(\mathbf{r}_{N_r} , l \right) \right]^\top \hat{\mathbf{p}} + \mathbf{z}(l) , \label{RR1}%
\end{align} 
where $G_r$ is the receiving antenna gain. 
Note that $\mathbf{y}(l)$ in (\ref{RR1}) is equal to the $l$-th column of $\mathbf{Y}$ in (\ref{sensing_channel}). 
Moreover, (\ref{lipp0})-(\ref{RR1}) unfold the EM propagation mechanism in (\ref{sensing_channel}).
\color{black}
According to (\ref{lipp0}), (\ref{lipp1}), and (\ref{RR1}), the EM property of the target is implicitly encoded in the received echo signals that are  transmitted through the sensing channel. 
Thus, we can leverage the estimated sensing channel matrix as the prior information to reconstruct the EM property of the target.

\begin{figure}[t]
  \centering  \centerline{\includegraphics[width = 8.4 cm]{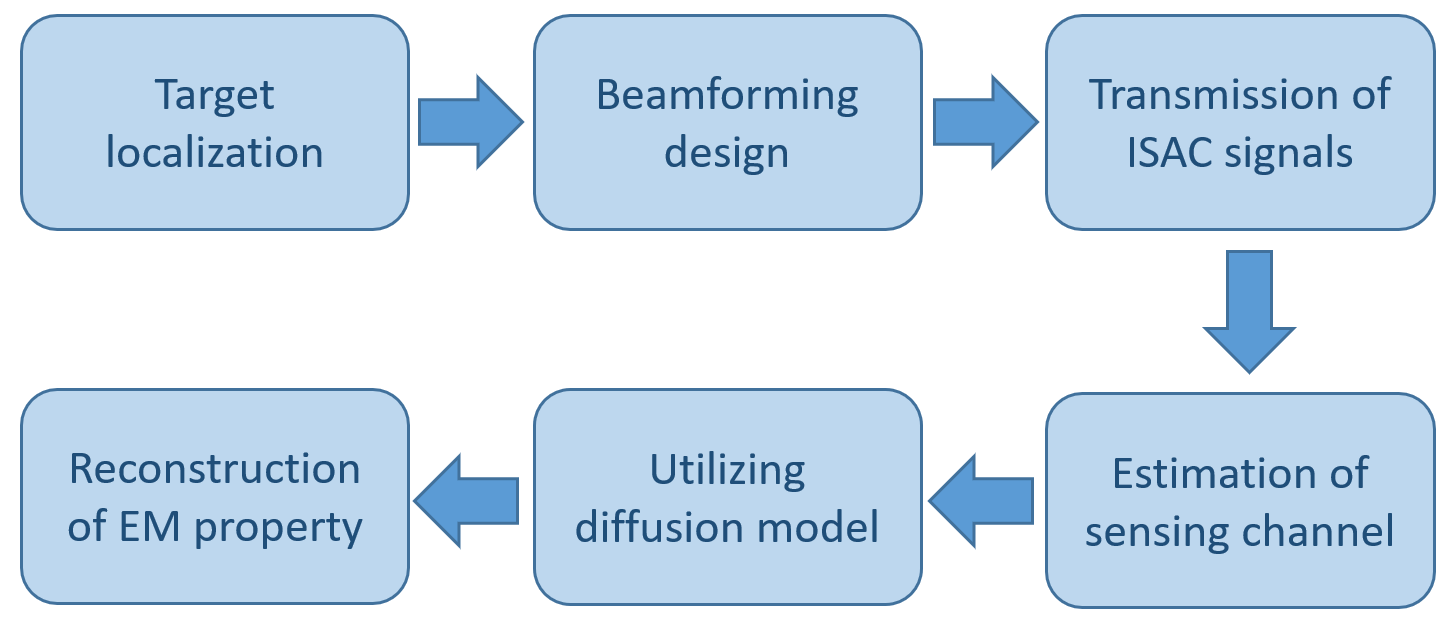}}
  \caption{\textcolor{blue}{The flowchart of the overall EM property sensing process.} 
  } 
  \label{flow_chart}
\end{figure}

\color{blue}
\theoremstyle{remark}
As the mapping from $\mathbf{x}(l)$ to $\mathbf{y}(l)$, the sensing channel $\mathbf{H}_s$ in (\ref{sensing_channel}) depends on the EM property of the target and is actually the composite mapping consisting of (\ref{h1}), (\ref{lipp0}), (\ref{lipp1}), and (\ref{RR1}). 
Specifically, (\ref{h1}) maps $\mathbf{x}(l)$ to $\mathbf{E}^i ({\mathbf {r}, l})$; 
(\ref{lipp0}) maps $\mathbf{E}^i ({\mathbf {r}, l})$ to $\mathbf{E}^t ({\mathbf {r}, l})$;
(\ref{lipp1}) maps $\mathbf{E}^t ({\mathbf {r}, l})$ to $\mathbf{E}^s ({\mathbf {r}, l})$;
(\ref{RR1}) maps $\mathbf{E}^s ({\mathbf {r}, l})$ to $\mathbf{y}(l)$. 
\color{black}
\color{blue}
Moreover, the overall EM property sensing process is summarized in Fig.~\ref{flow_chart}. 
Since the target localization method is well studied in ISAC, we do not discuss the localization method in this paper. 
\color{black}

\section{Diffusion Probabilistic Model for EM Property Sensing}
\textcolor{black}{
From the 3D formulation of EM scattering (\ref{lipp0})-(\ref{RR1}), we can sense the EM property of the target using inversion techniques \cite{li2018deepnis,mesh_free,zhang2022probabilistic}.
When employing the traditional mesh-based inversion techniques, the spatial arrangement of voxels is typically predetermined, which requires a comprehensive analysis of the entire domain of interest (DoI) to ascertain the EM property of the target.
The mesh-based approach, while thorough, incurs a significant computational cost due to the over-resolution of the background medium, which is not the focus of our interest.
Moreover, the results yielded by the 3D mesh-based inversion methods are not inherently conducive to direct visualization. 
}
On the other hand, the use of the point cloud methodology offers a more efficient and straightforward alternative to sense the EM property.
Point clouds inherently allow for the separation of the background medium from the target, which thus obviates the need to analyze the known background medium and substantially cuts down the computational overhead.
Moreover, the representation of data through the point cloud enables a clear and immediate visualization of the 3D target, which facilitates a more intuitive interpretation of the inversion results.

Define the $i$-th normalized non-dimensional 5D point $\mathbf{x}_i^{(0)} \in \mathbb{R}^{5 \times 1}$ that comprises both the 3D location information and the 2D EM property as 
\begin{align}
    \mathbf{x}_i^{(0)} = \left[\frac{x_i - x_c}{x_d}, \frac{y_i - y_c}{y_d}, \frac{ z_i - z_c }{ z_d }, 
    \frac{\epsilon_i}{\epsilon_0}, \frac{\sigma_i}{\epsilon_0 \omega}  \right]^\top , 
    \label{5D}
\end{align} 
where $x_i$, $y_i$, and $z_i$ denote the coordinates of the $i$-th point in each dimension; 
$x_c$, $y_c$, and $z_c$ denote the coordinates of the center of the target; 
$x_d$, $y_d$, and $z_d$ denote the characteristic length of each dimension that can be selected as the corresponding standard deviations without loss of generality.

Suppose the target can be represented by a 5D point cloud \( \mathcal{X}^{(0)} = \{\mathbf{x}_i^{(0)}\}_{i=1}^N \), consisting of \( N \) non-dimensional points with the corresponding EM property in (\ref{5D}). 
The 5D point cloud behaves like a collection of particles within an evolving thermodynamic system over time.
Each point \( \mathbf{x}_i^{(0)} \) in the point cloud is treated as being independently sampled from the point distribution, which will be denoted as \( q(\mathbf{x}_i^{(0)} | \hat{\mathbf{H}}_{s}) \). Here, \( \hat{\mathbf{H}}_{s} \) is considered as the prior condition that determines the distribution of $\mathbf{x}_i^{(0)}$.

\subsection{Forward Diffusion Process}
In the forward diffusion process, the original point cloud distribution progressively transitions into a noise distribution.
Let $\mathbf{x}_i^{(1: T)}$ denote the set of $\mathbf{x}_i^{(1)}, \cdots , \mathbf{x}_i^{(T)}$ that are sequentially generated from $\mathbf{x}_i^{(0)}$ with the maximum time step $T$. 
The forward diffusion process can be deliberately modeled as a Markov chain, i.e.,
\begin{align}
q\left(\mathbf{x}_i^{(1: T)} \mid \mathbf{x}_i^{(0)}\right)=\prod_{t=1}^T q\left(\mathbf{x}_i^{(t)} \mid \mathbf{x}_i^{(t-1)}\right),  
\label{forward_Markov}
\end{align}
where $q\left(\mathbf{x}_i^{(t)} \mid \mathbf{x}_i^{(t-1)}\right)$ represents the probability transition function, which simulates the evolution of point distributions from one time step to the next by incorporating noise into the current distribution.

Since each point in the point cloud is independently sampled from \( q(\mathbf{x}_i^{(0)} | \hat{\mathbf{H}}_{s}) \), 
the probability distribution of the overall point cloud is the product of the probability distribution of each point, i.e., 
\begin{align}
q\left(\mathcal{X}^{(1: T)} \mid \mathcal{X}^{(0)} \right) & = \prod_{i=1}^N q\left(\mathbf{x}_i^{(1: T)} \mid \mathbf{x}_i^{(0)}\right) .
\label{prod1}
\end{align}

The point distribution at each time step is assumed to follow a Gaussian distribution. 
Consequently, the transition probability function, which describes the noise addition in each time step, can be expressed as
\begin{align}
q\left(\mathbf{x}^{(t)}_{i} \mid \mathbf{x}^{(t-1)}_{i} \right)=\boldsymbol{\mathcal { N  }}\left(\mathbf{x}^{(t)}_{i} \mid \sqrt{1-\beta_t} \mathbf{x}^{(t-1)}_{i}, \beta_t \mathbf{I}\right), 
\label{beta0}
\end{align}
where $\beta_1, \cdots, \beta_T$ are linearly increasing hyperparameters that govern the diffusion intensity at each subsequent time step during the forward diffusion process. 

Referring to (\ref{beta0}), once the diffusion process between adjacent time steps is complete, 
it is possible to trace the transformation of the target from the initial state to any subsequent time step.
However, performing such iterative calculations for each time step during the training process is time-consuming and inefficient. 
By applying the Bayes' rule and using the renormalization technique, (\ref{beta0}) can be transformed to directly facilitate the calculation of the state of the point cloud at an arbitrary time step as
\color{blue}
\begin{align}
q\left(\mathbf{x}^{(t)}_{i} \mid \mathbf{x}^{(0)}_{i} \right) = \boldsymbol{\mathcal { N }}\left( \mathbf{x}^{(t)}_{i} \mid \sqrt{\bar{\alpha}_t} \mathbf{x}^{(0)}_{i},\left(1-\bar{\alpha}_t\right) \mathbf{I} \right) ,
\label{q0}
\end{align}
\color{black}
where $\alpha_t \triangleq 1 - \beta_t$ and $\bar{\alpha}_t \triangleq \prod_{p=1}^t \alpha_p$. 
The equation (\ref{q0}) provides the relationship between the intermediate state at any time step $t$ and the initial state, which significantly simplifies the calculation process. 

\begin{figure*}[t]
  \centering  \centerline{\includegraphics[width = 16.9cm]{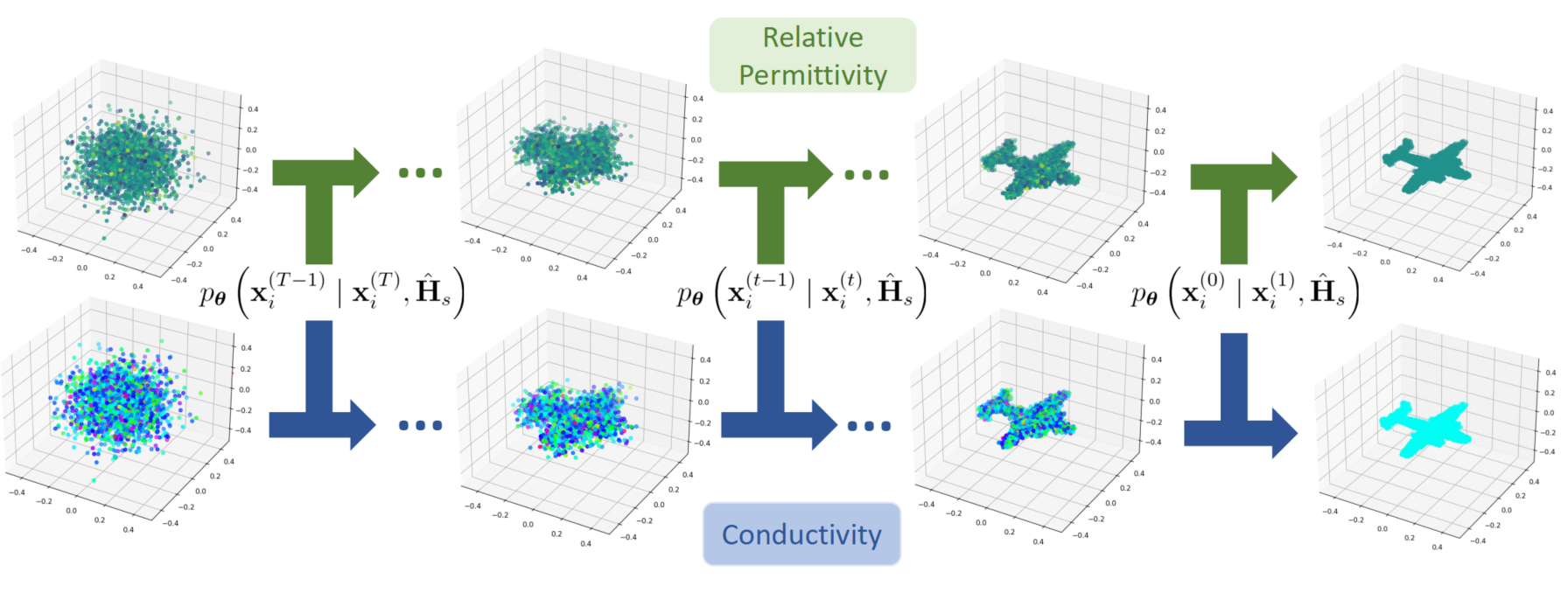}}
  \caption{The reverse diffusion process guided by the estimated sensing channel as the prior information. The shape, relative permittivity, and conductivity of the target are reconstructed simultaneously. \textcolor{blue}{The values of the relative permittivity are represented by the colors of the points in the first row, and the values of the conductivity are represented by the colors of the points in the second row.}}%
  \label{diffusion_process}%
\end{figure*}

\subsection{Reverse Diffusion Process} 
In this subsection, we aim to reconstruct the 5D point cloud representation of the target and the EM property of each point from the estimated sensing channel through the reverse diffusion process, which is analogous to the Langevin dynamics sampling procedures \cite{Langevin}.
In the reverse diffusion process, we can reformulate (\ref{q0}) as 
\begin{align}
\mathbf{x}^{(0)}_{i} = \frac{1}{\sqrt{\bar{\alpha}_t}}\left(\mathbf{x}^{(t)}_{i}-\sqrt{1-\bar{\alpha}_t} \boldsymbol{\epsilon}^{(t)}_i\right), \quad  \boldsymbol{\epsilon}^{(t)}_i \sim \mathcal{N}(\mathbf{0}, \mathbf{I}) .
\label{q00}
\end{align}
Thus, $\mathbf{x}^{(0)}_{i}$ can be estimated by denoising $\mathbf{x}^{(t)}_{i}$ through a sequential process.

Similar to the forward diffusion process (\ref{forward_Markov}), the reverse diffusion process can also be modeled as a Markov chain conditioned on the estimated sensing channel, i.e., 
\begin{align}
p_{{\boldsymbol{\theta}}}\left(\mathbf{x}_i^{(0: T)} \mid \hat{\mathbf{H}}_{s}\right)=p\left(\mathbf{x}_i^{(T)}\right) \prod_{t=1}^T p_{\boldsymbol{\theta}} \left(\mathbf{x}_i^{(t-1)} \mid \mathbf{x}_i^{(t)}, \hat{\mathbf{H}}_{s}\right) .
\label{p0}
\end{align} 

At the beginning of the reverse process, 
we start with $N$ points sampled from a predetermined Gaussian distribution $p\left(\mathbf{x}_i^{(T)}\right)$ that approximates $q\left(\mathbf{x}_i^{(T)}\right)$, which serves as the initial input.
These initial points are subsequently processed through the reverse Markov chain to ultimately delineate the original 5D point cloud.
Similar to (\ref{prod1}), there is
\begin{align}
p_{\boldsymbol{\theta}}\left(\mathcal{X}^{(0: T)} \mid \hat{\mathbf{H}}_{s}\right) & = \prod_{i=1}^N  p_{\boldsymbol{\theta}}\left(\mathbf{x}_i^{(0: T)} \mid \hat{\mathbf{H}}_{s}\right) .
\end{align}

In contrast to the learning-free forward diffusion process, the reverse diffusion process is learning-dependent. 
The diffusion model progressively retrieves the 5D point cloud constituting the target by sequentially denoising the initial Gaussian noise state. 
The shape, relative permittivity, and conductivity of the target are reconstructed simultaneously by the estimated sensing channel information as shown in Fig.~\ref{diffusion_process}.
Similarly to the forward diffusion process, the transition probability function in the reverse process can be articulated as
\begin{equation}
p_{\boldsymbol{\theta}}\left(\mathbf{x}_{i}^{(t-1)} \mid \mathbf{x}^{(t)}_{i}, \hat{\mathbf{H}}_{s}\right) \! = \!\boldsymbol{\mathcal { N }}\left(\mathbf{x}_{i}^{(t-1)} \!\mid \! \boldsymbol{\mu}_{{\boldsymbol{\theta}}}
\left(\mathbf{x}^{(t)}_{i}, t, \hat{\mathbf{H}}_{s}\right), \beta_t \mathbf{I}\right)\! ,
\label{p1}
\end{equation}
where $\boldsymbol{\mu}_{{\boldsymbol{\theta}}} \left(\mathbf{x}^{(t)}_{i}, t, \hat{\mathbf{H}}_{s}\right)$ is the estimated mean implemented by a neural network parameterized by ${\boldsymbol{\theta}}$. 
The covariance matrix of the added noise at time step \( t-1 \) is obtained based on the schedule in the forward diffusion process.
The corresponding estimated sensing channel \( \hat{\mathbf{H}}_{s} \), the point cloud representation of the target's EM property at time step \( t \), and the current time step \( t \) serve as conditions that enable the 
neural network to ascertain the mean of distribution at time step \( t-1 \). 
This, in turn, aids to determine the state of the point cloud at time step \( t-1 \). 
By reversing the forward diffusion process, we establish the precise mathematical expression of $q\left(\mathbf{x}_i^{(t-1)} \mid \mathbf{x}_i^{(t)}, \mathbf{x}_i^{(0)}\right)$ with the following closed-form Gaussian distribution
\begin{equation} 
q\left(\mathbf{x}_i^{(t-1)} \mid \mathbf{x}_i^{(t)}, \mathbf{x}_i^{(0)}\right)\!=\!\mathcal{N}\left(\mathbf{x}_i^{(t-1)} \mid \boldsymbol{\mu}_t\left(\mathbf{x}^{(t)}_{i}, \mathbf{x}^{(0)}_{i}\right), \gamma_t \mathbf{I} 
 \right)\!, 
\label{q1} 
\end{equation}
where the mean and the covariance are calculated according to the Bayes' rule as \cite{VLB_diffusion}
\begin{align} 
\boldsymbol{\mu}_t\left(\mathbf{x}^{(t)}_{i}, \mathbf{x}^{(0)}_{i}\right) & =\frac{\sqrt{\bar{\alpha}_{t-1}} \beta_t}{1-\bar{\alpha}_t} \mathbf{x}^{(0)}_{i}+\frac{\sqrt{\alpha_t}\left(1-\bar{\alpha}_{t-1}\right)}{1-\bar{\alpha}_t} \mathbf{x}^{(t)}_{i} , \label{mu} \\
\gamma_t & =\frac{1-\bar{\alpha}_{t-1}}{1-\bar{\alpha}_t} \beta_t . 
\label{gamma2}
\end{align}
Substituting (\ref{q00}) into (\ref{mu}), there is 
\begin{align} 
\boldsymbol{\mu}_t\left(\mathbf{x}^{(t)}_{i}, \mathbf{x}^{(0)}_{i}\right) & \! = \! \frac{1}{\sqrt{\alpha_t}}\left(\mathbf{x}^{(t)}_{i}-\frac{\beta_t}{\sqrt{1-\bar{\alpha}_t}} \boldsymbol{\epsilon}^{(t)}_i\right) \!,  \boldsymbol{\epsilon}^{(t)}_i \sim \mathcal{N}(\mathbf{0}, \mathbf{I}) .
\label{mu2}
\end{align} 
The procedure to estimate the 5D point cloud \(\hat{\mathcal{X}}^{(0)}\) is summarized in Algorithm~\ref{algs}.

\begin{algorithm}[t]
\caption{5D Point Cloud Estimation} 
\begin{algorithmic}[1]
    \Require Estimated sensing channel \(\hat{\mathbf{H}}_{s}\) corresponding to \(\hat{\mathcal{X}}^{(0)}\), $p_{{\boldsymbol{\theta}}}(\cdot | \cdot ,  \cdot)$, and $T$
    \State Sample \(\hat{\mathcal{X}}^{(T)} \sim \mathcal{N}(\mathbf{0}, \mathbf{I})\)
    \For{$t = T, \cdots, 1$} 
        \State Sample \(\hat{\mathcal{X}}^{(t-1)} \sim p_{{\boldsymbol{\theta}}}(\hat{\mathcal{X}}^{(t-1)} | \hat{\mathcal{X}}^{(t)}, \hat{\mathbf{H}}_{s})\)
    \EndFor \\
\Return \(\hat{\mathcal{X}}^{(0)}\)
\end{algorithmic}
\label{algs}
\end{algorithm} 

\subsection{Variational Lower Bound Maximization} 
The training goal in the reverse diffusion process is to maximize the log-likelihood function expectation of the 5D point cloud with EM property, expressed as \( \mathbb{E}_{q\left(\mathbf{x}^{(0)}_i\right)}[\ln p_{\boldsymbol{\theta}} (\mathbf{x}^{(0)}_{i} \mid \hat{\mathbf{H}}_{s} )] \).
However, direct optimization of the log-likelihood function expectation is impractical. 
Instead, we can maximize the variational lower bound (VLB)\footnote{VLB is also known as the evidence lower bound (ELBO) in some literature.} of $\mathbb{E}_{q\left(\mathbf{x}^{(0)}_i\right)}[\ln p_{\boldsymbol{\theta}} (\mathbf{x}^{(0)}_{i} \mid \hat{\mathbf{H}}_{s} )]$, derived as 
\begin{align}
& \mathbb{E}_{q\left(\mathbf{x}^{(0 )}_i\right)} \ln p_{\boldsymbol{\theta}}\left(\mathbf{x}^{(0)}_{i}  \mid \hat{\mathbf{H}}_{s} \right) \nonumber\\
& = \mathbb{E}_{q\left(\mathbf{x}^{(0 )}_i\right)} \ln \left[ \mathbb{E}_{q\left(\mathbf{x}^{(1: T)}_i \mid \mathbf{x}^{(0)}_{i}\right)}
\frac{p_{\boldsymbol{\theta}} \left(\mathbf{x}^{(0)}_{i}  \mid \hat{\mathbf{H}}_{s} \right)}{q\left(\mathbf{x}^{(1: T)}_i \mid \mathbf{x}^{(0)}_{i}\right)}    \right] 
\nonumber \\ 
& \overset{(a)}{\geq} \mathbb{E}_{q\left(\mathbf{x}^{(0 )}_i\right)} \mathbb{E}_{q\left(\mathbf{x}^{(1: T)}_i \mid \mathbf{x}^{(0)}_{i}\right)} \ln \left[ 
\frac{p_{\boldsymbol{\theta}} \left(\mathbf{x}^{(0)}_{i}  \mid \hat{\mathbf{H}}_{s} \right)}{q\left(\mathbf{x}^{(1: T)}_i \mid \mathbf{x}^{(0)}_{i}\right)}    \right] 
\nonumber \\ 
& = \mathbb{E}_{q\left(\mathbf{x}^{(0: T)}_i\right)} \ln \left[ \frac{p_{\boldsymbol{\theta}}\left(\mathbf{x}^{(0: T)}_i  \mid \hat{\mathbf{H}}_{s} \right)}{q\left(\mathbf{x}^{(1: T)}_i \mid \mathbf{x}^{(0)}_{i}\right)  }\right] \triangleq \mathrm{VLB}(\hat{\mathbf{H}}_{s}) ,
\label{VLB0}
\end{align}
where $ \overset{(a)}{\geq}$ comes from the Jensen inequality for the log function. 
In order to clarify the logical structure of $\mathrm{VLB}(\hat{\mathbf{H}}_{s})$, we further break down $\mathrm{VLB}(\hat{\mathbf{H}}_{s})$ into its three constituent components by substituting (\ref{p0}) into (\ref{VLB0}) as
\begin{align} 
& \mathrm{VLB}(\hat{\mathbf{H}}_{s}) \! = \!   \mathbb{E}_{q\left( \mathbf{x}^{(0: T)}_i \right)} \left[ \underbrace{ \!\ln p_{\boldsymbol{\theta}}\left(\mathbf{x}^{(0)}_{i} \! \mid \!\mathbf{x}^{(1)}_{i}  , \hat{\mathbf{H}}_{s} \right) }_{L_1}
\right.  \nonumber \\ 
& + \underbrace{D_{\mathrm{KL}}\left(
q\left(\mathbf{x}^{(T)}_i \mid \mathbf{x}^{(0)}_{i}\right) \| p \left(\mathbf{x}^{(T)}_i\right) \right) }_{L_2} 
   \nonumber \\ 
& \left. + \sum_{t=2}^T \underbrace{D_{\mathrm{KL}} \! \left(q\left(\mathbf{x}^{(t-1)}_{i} \mid \mathbf{x}^{(t)}_{i}, \mathbf{x}^{(0)}_{i}\right) \! \| p_{\boldsymbol{\theta}}\left(\mathbf{x}^{(t-1)}_{i} \mid \mathbf{x}^{(t)}_{i} , \hat{\mathbf{H}}_{s} \right)\right)}_{\text{diffusion loss}}\right] \!  \! ,
\label{VLB2}
\end{align} 
where $D_{\mathrm{KL}}(\cdot \| \cdot)$ denotes the Kullback-Leibler (KL) divergence. 
Note that for a long time diffusion sequence with a relatively large $T$, $L_1$ is relatively small compared to the diffusion loss. 
Besides, $L_2$ is the KL divergence of two deterministic Gaussian distributions irrelevent to $\boldsymbol{\theta}$ and is thus a constant.
Therefore, we aim to minimize the diffusion loss that is the sum of a series of KL divergences. 
Since both $q\left(\mathbf{x}^{(t-1)}_i \mid \mathbf{x}^{(t)}_{i}, \mathbf{x}^{(0)}_{i}\right)$ and $p_{\boldsymbol{\theta}}\left(\mathbf{x}^{(t-1)}_{i} \mid \mathbf{x}^{(t)}_{i}, \hat{\mathbf{H}}_{s}\right)$ are Gaussian distributions according to (\ref{p1}) and (\ref{q1}),  
their KL divergence at time step $t$ can be computed as 
\begin{align} 
& D_{\mathrm{KL}}\left(q\left(\mathbf{x}^{(t-1)}_{i} \mid \mathbf{x}^{(t)}_{i}, \mathbf{x}^{(0)}_{i}\right) \left\| \right. p_{\boldsymbol{\theta}}\left(\mathbf{x}^{(t-1)}_{i} \mid \mathbf{x}^{(t)}_{i}\right)\right) 
= \frac{1}{2} \left[ 5 \frac{\gamma_t}{\beta_t} \right. \nonumber\\ 
& \left.+ \frac{1}{\beta_t} \left 
\|\boldsymbol{\mu}_{{\boldsymbol{\theta}}}
\left(\mathbf{x}^{(t)}_{i}, t, \hat{\mathbf{H}}_{s}\right) - \boldsymbol{\mu}_t\left(\mathbf{x}^{(t)}_{i}, \mathbf{x}^{(0)}_{i}\right) \right \|_2^2 - 5 + 5 \ln\frac{\beta_t}{\gamma_t} \right] \nonumber\\
& \overset{(a)}{=} \frac{1}{2} \! \! \left[ 5 \frac{1-\bar{\alpha}_{t-1}}{1-\bar{\alpha}_t} - 5 - 5 \ln\left(\frac{1-\bar{\alpha}_{t-1}}{1-\bar{\alpha}_t}\right)  \right. \nonumber\\
& \left. + \frac{1}{\beta_t} \left 
\|\boldsymbol{\mu}_{{\boldsymbol{\theta}}}
\left(\mathbf{x}^{(t)}_{i}, t, \hat{\mathbf{H}}_{s}\right) \! - \! \frac{1}{\sqrt{\alpha_t}}\left(\mathbf{x}^{(t)}_{i}-\frac{\beta_t}{\sqrt{1-\bar{\alpha}_t}} \boldsymbol{\epsilon}^{(t)}_i\right)  \right \|_2^2  \right] , 
\label{KL1}
\end{align} 
where $\overset{(a)}{=}$ is derived from (\ref{gamma2}) and (\ref{mu2}). 
In order to maximize $\mathrm{VLB}(\hat{\mathbf{H}}_{s})$, we aim to minimize the only variable  $\left 
\|\boldsymbol{\mu}_{{\boldsymbol{\theta}}}
\left(\mathbf{x}^{(t)}_{i}, t, \hat{\mathbf{H}}_{s}\right) \! - \! \frac{1}{\sqrt{\alpha_t}}\left(\mathbf{x}^{(t)}_{i}-\frac{\beta_t}{\sqrt{1-\bar{\alpha}_t}} \boldsymbol{\epsilon}^{(t)}_i\right) \right \|_2^2$ in (\ref{KL1}). 
Let $\boldsymbol{\epsilon}_{{\boldsymbol{\theta}}}\left(\mathbf{x}^{(t)}_{i}, t, \hat{\mathbf{H}}_{s}\right)$ denote the reverse noise estimator for the 
normalized noise added to $\mathbf{x}^{(t - 1)}_{i} $ in (\ref{mu2}) during the forward diffusion process. Then, there is 
\begin{align}
& \boldsymbol{\mu}_{{\boldsymbol{\theta}}}\left(\mathbf{x}^{(t)}_{i}, t, \hat{\mathbf{H}}_{s}\right)=\frac{1}{\sqrt{\alpha_t}}\left[\mathbf{x}^{(t)}_{i}-\frac{\beta_t}{\sqrt{1-\bar{\alpha}_t}} \boldsymbol{\epsilon}_{{\boldsymbol{\theta}}}\left(\mathbf{x}^{(t)}_{i}, t, \hat{\mathbf{H}}_{s}\right)\right] .
\label{KL2} 
\end{align} 
According to (\ref{KL2}), minimizing $\left 
\|\boldsymbol{\mu}_{{\boldsymbol{\theta}}}
\left(\mathbf{x}^{(t)}_{i}, t, \hat{\mathbf{H}}_{s}\right) \! - \! \frac{1}{\sqrt{\alpha_t}}\left(\mathbf{x}^{(t)}_{i}-\frac{\beta_t}{\sqrt{1-\bar{\alpha}_t}} \boldsymbol{\epsilon}^{(t)}_i\right)  \right \|_2^2$
is thereby equivalent to minimizing $\left\|\boldsymbol{\epsilon}^{(t)}_i-\boldsymbol{\epsilon}_{{\boldsymbol{\theta}}}\left(\mathbf{x}^{(t)}_{i}, t, \hat{\mathbf{H}}_{s} \right)\right\|_2^2$.
Since the reverse noise estimator is hard to formulate explicitly, we propose to model it as a neural network.  
To train this neural network, the loss function of the reverse noise estimator can be formulated by substituting (\ref{KL2}) into (\ref{KL1}) and dropping the constants, i.e.,
\begin{align}
 \text {Loss}_1 & = \mathbb{E}_{t \sim \mathcal{U}\{1, 2, \dots, T\} , \mathbf{x}^{(0)}_{i} \sim q\left(\mathbf{x}^{(0)}_{i}\right) , \boldsymbol{\epsilon}^{(t)}_i \sim \mathcal{N}(\mathbf{0},\mathbf{I})}\left[ 
    \frac{\beta_t}{2 \alpha_t (1-\bar{\alpha}_t)} \right. \nonumber\\
    & \times \left\|  \left.  
 \boldsymbol{\epsilon}^{(t)}_i-\boldsymbol{\epsilon}_{{\boldsymbol{\theta}}}\left(\mathbf{x}^{(t)}_{i}, t, \hat{\mathbf{H}}_{s} \right)\right\|_2^2\right] \nonumber\\
 & \!\!\!\!\!\!\!\!\!\!\!\!  \overset{(a)}{\approx} \frac{1}{T N} \sum_{t=1}^{T} \sum_{i=1}^{N}  \frac{\beta_t}{2 \alpha_t (1-\bar{\alpha}_t)} \left\|\boldsymbol{\epsilon}^{(t)}_i-\boldsymbol{\epsilon}_{{\boldsymbol{\theta}}}\left(\mathbf{x}^{(t)}_{i}, t, \hat{\mathbf{H}}_{s} \right)\right\|_2^2 \! , 
 \label{loss}
\end{align} 
\color{blue}
where $\mathcal{U}\{1, 2, \dots, T\}$ denotes the discrete uniform distribution, \color{black} and 
$\overset{(a)}{\approx}$ approximates the loss function by the weighted 
mean squared error between the added noise and the estimated noise. 

\color{blue}
In practical training process, we replace $\hat{\mathbf{H}}_{s}$ in $\text {Loss}_1$ with $\mathbf{H}_{s}$ and thus aim to maximize $\mathrm{VLB}(\mathbf{H}_{s})$ instead of $\mathrm{VLB}(\hat{\mathbf{H}}_{s})$.
Given that $\hat{\mathbf{H}}_{s}$ is corrupted with noise in the practical ISAC process, we need to derive the difference between the VLBs evaluated with the true sensing channel $\mathbf{H}_s$ and the noisy estimate $\hat{\mathbf{H}}_s$.
According to the definition of VLB in (\ref{VLB0}), there is
\begin{align}
& \Delta \mathrm{VLB} \! \triangleq \! \mathrm{VLB}(\mathbf{H}_s) \!-\! \mathrm{VLB}(\hat{\mathbf{H}}_s) \!=\! \mathbb{E}_{q(\mathbf{x}_i^{(0:T)})}\! \left[ \ln\frac{ p_{\boldsymbol{\theta}}(\mathbf{x}_i^{(0:T)} \mid \mathbf{H}_s) }{ p_{\boldsymbol{\theta}}(\mathbf{x}_i^{(0:T)} \mid \hat{\mathbf{H}}_s)} \right] \nonumber \\ 
& = \mathbb{E}_{q(\mathbf{x}_i^{(0:T)})} \left[ \sum_{t=1}^T \ln\frac{ p_{\boldsymbol{\theta}} \left(\mathbf{x}^{(t-1)}_i \mid \mathbf{x}^{(t)}_i, \mathbf{H}_s\right)}{ p_{\boldsymbol{\theta}} \left(\mathbf{x}^{(t-1)}_i \mid \mathbf{x}^{(t)}_i, \hat{\mathbf{H}}_s\right)} \right].
\end{align}
When the diffusion model is well trained, $p_{\boldsymbol{\theta}}(\mathbf{x}_i^{(0:T)}\mid \mathbf{H}_s)$ is very close to $q(\mathbf{x}_i^{(0:T)})$. 
Thus, $\Delta \mathrm{VLB}$ can be approximated by the sum of KL divergence when the expectation is taken with respect to distribution $p_{\boldsymbol{\theta}} \left(\mathbf{x}^{(t-1)}_i \mid \mathbf{x}^{(t)}_i, \mathbf{H}_s\right)$ as the approximation of $q\left(\mathbf{x}^{(t-1)}_i \mid \mathbf{x}^{(t)}_i \right)$. 
Then, there is 
\begin{align}
& \Delta \mathrm{VLB} \approx
\left. \mathbb{E}_{q(\mathbf{x}_i^{(0:T)})}\left\{ \sum_{t=1}^T D_{\mathrm{KL}} \left( p_{\boldsymbol{\theta}} \left(\mathbf{x}^{(t-1)}_i \mid \mathbf{x}^{(t)}_i, \mathbf{H}_s\right) \right \|   \right. \right.
\nonumber \\
& \left. \left. p_{\boldsymbol{\theta}} \left(\mathbf{x}^{(t-1)}_i \mid \mathbf{x}^{(t)}_i, \hat{\mathbf{H}}_s\right) \right)  \right\}
\nonumber \\ 
& = \mathbb{E}_{q(\mathbf{x}_i^{(0:T)})}\left\{ \sum_{t=1}^T \frac{1}{2 \beta_t} \left\| \boldsymbol{\mu}_{\boldsymbol{\theta}}(\mathbf{x}^{(t)}_i, t, \mathbf{H}_s) - \boldsymbol{\mu}_{\boldsymbol{\theta}}(\mathbf{x}^{(t)}_i, t, \hat{\mathbf{H}}_s) \right\|_2^2 \right\}.
\end{align} 

Denote $\mathbf{J}_{\boldsymbol{\mu}}(\mathbf{x}^{(t)}_i, t, \mathbf{H}_s)$ as the Jacobian matrix of $\boldsymbol{\mu}_{\boldsymbol{\theta}}(\mathbf{x}^{(t)}_i, t, \mathbf{H}_s)$ with respect to $\mathbf{H}_s$.
Denote the distribution of $\hat{{\mathbf{H}}}_{s}$ as $f(\hat{{\mathbf{H}}}_{s})$. 
Since there is $\text{vec}(\hat{{\mathbf{H}}}_{s}) \sim \mathcal{C N}\left(\text{vec}({{\mathbf{H}}}_{s}), \sigma_s^2\left(\mathbf{X}^*  \mathbf{X}^\top \otimes \mathbf{I}_{N_r}\right)^{-1}\right)$, we obtain
\begin{align}
& \mathbb{E}_{f(\hat{{\mathbf{H}}}_{s})} (\Delta \mathrm{VLB})  = \frac{\sigma_s^2}{2} \mathbb{E}_{q(\mathbf{x}_i^{(0:T)})}\left\{ \sum_{t=1}^T \frac{1}{\beta_t} \times \right. \nonumber\\
& \text{tr}\left[ \mathbf{J}_{\boldsymbol{\mu}}(\mathbf{x}^{(t)}_i, t, \mathbf{H}_s)^H \mathbf{J}_{\boldsymbol{\mu}}(\mathbf{x}^{(t)}_i, t, \mathbf{H}_s)  \left.  (\mathbf{X}^* \mathbf{X}^\top \otimes \mathbf{I}_{N_r})^{-1} \right] \right\}.
\label{insight}
\end{align}

\begin{remark}
%
Equation (\ref{insight}) presents the insight into  how the error of sensing channel estimation degrades the performance of EM property sensing.
When the power of noise at the receiver $\sigma_s^2$ becomes larger, the error of sensing channel estimation becomes larger. 
Then, $\mathbb{E}_{f(\hat{{\mathbf{H}}}_{s})} (\Delta \mathrm{VLB})$ becomes larger, and $\mathbb{E}_{f(\hat{{\mathbf{H}}}_{s})} (\mathrm{VLB}(\hat{\mathbf{H}}_s))$ becomes consequently smaller.
As a result, the EM property sensing performance worsens with a relatively low $\mathbb{E}_{f(\hat{{\mathbf{H}}}_{s})} (\mathrm{VLB}(\hat{\mathbf{H}}_s))$. 
Improving the accuracy of channel estimation is thus important to enhance the performance of EM property sensing. 
If $\mathbf{J}_{\boldsymbol{\mu}}(\mathbf{x}^{(t)}_i, t, \mathbf{H}_s)^H \mathbf{J}_{\boldsymbol{\mu}}(\mathbf{x}^{(t)}_i, t, \mathbf{H}_s)$ is an identity matrix as a special case, then minimizing $\mathbb{E}_{f(\hat{{\mathbf{H}}}_{s})} (\Delta \mathrm{VLB})$ is equivalent to minimizing the channel estimation error with a well trained diffusion model.
\end{remark}

\color{black}

\subsection{Reverse Noise Estimator}
\begin{figure*}[t]
  \centering
\centerline{\includegraphics[width=16.9cm]{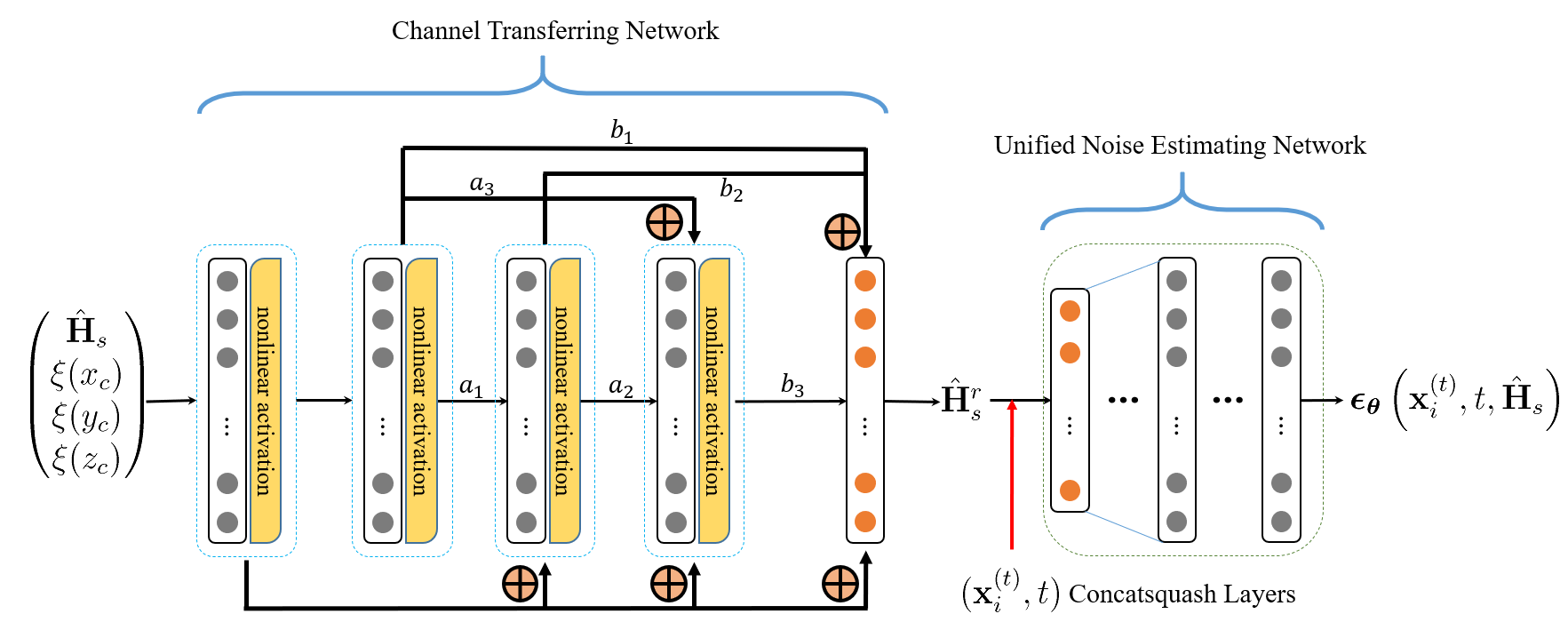}}
  \caption{Schematic diagram of the reverse noise estimator, which is composed of two cascaded neural networks: the channel transferring network and the unified noise estimating network.}
  \label{noise_estimator}
\end{figure*}
Since the sensing channel is influenced by not only the EM property but also the location of the target, 
training the reverse noise estimator is relatively easy when the target is situated at a predefined reference location.
Define $\mathbf{H}_s^r$ as the reference sensing channel when the target is translated to the reference location $(x_r,y_r,z_r)$ with the shape, posture, and EM property unchanged.
We will first transfer the sensing channel to the reference sensing channel and then estimate the noise in the diffusion process.
The reverse noise estimator $\boldsymbol{\epsilon}_{{\boldsymbol{\theta}}}\left(\mathbf{x}_i^{(t)}, t, \hat{\mathbf{H}}_s\right)$ is composed of two cascaded neural networks: the channel transferring network and the unified noise estimating network, as shown in Fig.~\ref{noise_estimator}.

The channel transferring network plays the role of estimating the reference sensing channel $\hat{\mathbf{H}}_s^r$ by taking $\hat{\mathbf{H}}_s$ and $(x_c,y_c,z_c)$ as the inputs. 
Despite the fact that neural networks can be universal function approximators, we found that having the channel transferring network directly operate on the input coordinates $(x_c,y_c,z_c)$ leads to relatively poor representation of the fine-grained spatial variations in location. This is consistent with recent work by \cite{rahaman2019spectral}, which shows that the deep neural networks are biased towards learning lower-frequency functions. 
\textcolor{black}{
However, by mapping the inputs to a higher-dimensional space with high-frequency sinusoidal functions prior to their entry into the channel transferring network, it becomes possible to achieve a more accurate fit of data with complex spatial intricacies. 
}

Define the positional encoding function $\xi(\cdot)$ as a mapping from $\mathbb{R}$ into a higher-dimensional space $\mathbb{R}^{ (2 \bar{L}+1)  \times 1 }$.
Then we apply $\xi(\cdot)$ separately to each of the three coordinate values as $[\xi(x_c) , \xi(y_c) , \xi(z_c)]$.  
The positional encoding function can be defined as
\begin{align} 
\xi(p) & = \left[ p , \sin \left(2^0 \pi p\right), \cos \left(2^0 \pi p\right), \cdots, \right. \nonumber \\
& \left. \sin \left(2^{\bar{L}-1} \pi p\right), \cos \left(2^{\bar{L}-1} \pi p\right) \right] ^ \top, p \in \{ x_c, y_c , z_c\} .  
\label{position_encode}
\end{align} 
\textcolor{black}{
The channel transferring network takes $\xi(x_c) , \xi(y_c) , \xi(z_c)$, and $\hat{\mathbf{H}}_s$ as the inputs that are concatenated into a real-valued vector $ \left[ \text{vec}(\Re(\hat{\mathbf{H}}_s))^\top, \text{vec}(\Im(\hat{\mathbf{H}}_s))^\top , \xi(x_c)^\top , \xi(y_c)^\top , \xi(z_c)^\top \right]^\top$ for further processing. 
} %
The reference sensing channel is estimated by the regular Neural ODE \cite{finlay2020train,lin2021deep} with DenseNet structure \cite{densenet}.
We set the dimensions of all the hidden layers as $512$. 
The parameters $(a_1,a_2,a_3,b_1,b_2,b_3)$ in Fig.~\ref{noise_estimator} denote the trainable coefficients multiplied to the output of the previous layer.

In order to train the channel transferring network, we define the loss function as the normalized mean square error (NMSE) of the reference sensing channel estimation, i.e.,
\begin{align} 
\text {Loss}_2 = \mathrm{NMSE}(\hat{\mathbf{H}}_s^r) =
\frac{\left\|\mathbf{H}_s^r-\hat{\mathbf{H}}_s^r \right\|_F^{2}}{\left\|\mathbf{H}_s^r \right\|_F^{2}} .
\label{loss2}
\end{align} 

As the second part of the reverse noise estimator, 
the unified noise estimating network is composed of a series of $L_{cs}$ concatsquash layers \cite{grathwohl2018ffjord} defined as 
\begin{align}
\mathbf{e}^{n+1} = \left(\mathbf{W}^{n}_1 \mathbf{e}^{n}+\mathbf{b}^{n}_1\right)
\odot \text{Sigmoid}\left(\mathbf{W}^{n}_2 \mathbf{ c } + \mathbf{b}^{n}_2\right) + \mathbf{W}^{n}_3 \mathbf{c},  
\end{align}
where $\mathbf{e}^{n}$ is the input to the $n$-th layer and $\mathbf{e}^{n+1}$ is the output.
\textcolor{blue}{%
Besides, $\text{Sigmoid}(\mathbf{ a }) = \frac{1}{1 + e^{-\mathbf{a}}}$ is the Sigmoid function which is applied to a vector in an element-wise manner.} 
The input to the first layer is the 5D point $\mathbf{x}_i^{(t)}$, and the output of the last layer is the estimated noise  $\boldsymbol{\epsilon}_{{\boldsymbol{\theta}}}\left(\mathbf{x}^{(t)}_{i}, t, \hat{\mathbf{H}}_{s} \right)$. 
Moreover, we define the real-valued concatenated context vector as $\mathbf{c}=\left[t, \sin (t), \cos (t), \sin (2t), \cos (2t), \!\text{vec}(\Re(\hat{\mathbf{H}}_s^r))^\top\!, \!\! \text{vec}(\Im(\hat{\mathbf{H}}_s^r))^\top \!\! \right] ^\top\!\!\!\!.$
\textcolor{black}{
The context vector incorporates the embedded random Fourier features of the time step $t$ \cite{Fourier_feature} and the estimated reference sensing channel matrix $\hat{\mathbf{H}}_s^r$. }%
Besides, $\{\mathbf{W}^{n}_1, \mathbf{W}^{n}_2, \mathbf{W}^{n}_3, \mathbf{b}^{n}_1, \mathbf{b}^{n}_2 \}_{n=1:10}$ are trainable weights and biases in the neural network. 
We use the Swish activation function to create non-linearity between the layers \cite{swish}. 

In the training procedure, we first train the unified noise estimating network with $\text{Loss}_1$ 
and then train the channel transferring network with $\text{Loss}_2$, where we assume the center of the target is sampled from a predetermined sensing area. 
Since the two loss functions are irrelevant to each other, the two networks are trained independently. 
With the loss function (\ref{loss}) and (\ref{loss2}), the procedure to train the reverse noise estimator 
$\boldsymbol{\epsilon}_{{\boldsymbol{\theta}}}\left(\mathbf{x}^{(t)}_{i}, t, \hat{\mathbf{H}}_{s} \right)$ is summarized in Algorithm~\ref{algt}. 

\begin{algorithm}[t]
\caption{Training Reverse Noise Estimator } 
\begin{algorithmic}[1]
\Repeat
    \State Sample \(\mathcal{X}^{(0)} \sim q (\mathcal{X}^{(0)})\)
    \State Select the reference sensing channel \( \mathbf{H}_{s}^r \) corresponding \Statex \ \ \ \ \ to \(\mathcal{X}^{(0)}\)
    \For{ $t = 1, \cdots, T$  }
    \For{ $i = 1, \cdots, N$  }
    \State Sample \(\boldsymbol{\epsilon}^{(t)}_i \sim \mathcal{N}(\mathbf{0}, \mathbf{I})\)
    \State Compute \(\mathbf{x}_i^{(t)} \sim q(\mathbf{x}_i^{(t)} | \mathbf{x}_i^{(0)})\)
    \EndFor
    \EndFor
    \State Compute \( \text {Loss}_1\) according to (\ref{loss})
    \State Compute \(\nabla\text {Loss}_1\), and perform gradient descent. 
\Until converged 
\Repeat 
    \State Sample \(\mathcal{X}^{(0)} \sim q (\mathcal{X}^{(0)})\) and sample $(x_c,y_c,z_c)$ \Statex  \ \ \ \ \  uniformly and randomly in the sensing area  
    \State Select \( \mathbf{H}_{s} \) and \( \mathbf{H}_{s}^r \) corresponding to \(\mathcal{X}^{(0)}\) and \Statex  \ \ \ \ \ $(x_c,y_c,z_c)$ 
    \State Compute $\hat{\mathbf{H}}_{s}^r$ and compute \( \text {Loss}_2\) according to (\ref{loss2}) 
    \State Compute \(\nabla\text {Loss}_2\), and perform gradient descent.
\Until converged 
\end{algorithmic}
\label{algt}
\end{algorithm}


\section{Beamforming Design with Tradeoff Between Sensing and Communications}
Since it is relatively hard to directly minimize the Chamfer distance between the reconstructed 5D point cloud and the ground truth through the beamforming design, 
we aim to minimize the CRB of sensing channel estimation in (\ref{crb2}) alternatively, which is positively related to the discrepancy of 5D point cloud reconstruction. 
In particular, we would optimize the transmit covariance matrix $\mathbf{S}_x$ to minimize the CRB of sensing channel estimation, while guaranteeing the minimum communications rate $\bar{R}$ for each UE, and subject to the maximum transmit power constraint $P$. 
Specifically, the communications achievable rate-constrained CRB minimization problem is formulated as 
\begin{subequations}
\begin{align}
 \text { (P1) : } & \min _{\mathbf{S}_x, \mathbf{R}_{k,k=1,\cdots,K }} \operatorname{tr}(\textrm{CRB})= \frac{N_r \sigma_s^2}{L} \operatorname{tr}\left(\mathbf{S}_x^{-1}\right) \\
 \text { s.t. } & \min _{k} \log_2\left(1 + \frac{\mathbf{h}_k^H \mathbf{R}_k \mathbf{h}_k }{\sigma_k^2 + \mathbf{h}_k^H (\mathbf{S}_x - \mathbf{R}_k) 
\mathbf{h}_k} \right) \geq \bar{R}, \\
& \operatorname{tr}\left(\mathbf{S}_x\right) \leq P , \\
& \mathbf{S}_x \succeq \mathbf{0}, \quad \mathbf{R}_k \succeq 0 , \quad k=1, \cdots, K, \\
& \operatorname{rank}\left(\mathbf{R}_k\right)=1, \quad k=1, \cdots, K,  \\
& \mathbf{R}-\sum_{k=1}^K \mathbf{R}_k \succeq 0, \quad k=1, \cdots, K
.
\end{align}
\end{subequations}

By introducing $\Gamma = \frac{1}{2^{\bar{R}}-1}$, problem (P1) can be equivalently reformulated as
\begin{subequations}
\begin{align} 
\text { (P2) : } & \min _{\mathbf{S}_x, \mathbf{R}_{k, k = 1, \cdots , K } }  \operatorname{tr} \left(\mathbf{S}_x^{-1}\right) \\
\text { s.t. } \quad & (1 + \Gamma) \mathbf{h}_k^H \mathbf{R}_k \mathbf{h}_k \geq  \sigma_k^2 + \mathbf{h}_k^H \mathbf{S}_x \mathbf{h}_k, \quad k=1, \cdots, K , \label{36b} \\ 
& \operatorname{tr}\left(\mathbf{S}_x\right) \leq P , \label{36c} \\
& \mathbf{S}_x \succeq \mathbf{0}, \quad \mathbf{R}_k \succeq 0 , \quad k=1, \cdots, K, \label{36d}\\
& \operatorname{rank}\left(\mathbf{R}_k\right)=1, \quad k=1, \cdots, K, \label{rank1} \\
& \mathbf{S}_x-\sum_{k=1}^K \mathbf{R}_k \succeq 0, \quad k=1, \cdots, K.  \label{36f}  
\end{align}  
\end{subequations}  

By dropping the rank-one constraints (\ref{rank1}), (P2) becomes a convex optimization problem and can be solved via CVX efficiently \cite{cvx}. 
Denote $\tilde{\mathbf{S}}_x$ and $\tilde{\mathbf{R}}_k$ as the feasible solutions to problem (P2) without the rank-one constraints (\ref{rank1}), which are referred to as the semidefinite relaxation (SDR) solutions \cite{SDR}. 
We can then construct the communications beamforming matrix $\hat{\mathbf{W}}_{\mathrm{c}}=\left[\hat{\mathbf{w}}_{\mathrm{c}, 1}, \hat{\mathbf{w}}_{\mathrm{c}, 2}, \cdots, \hat{\mathbf{w}}_{\mathrm{c}, K}\right]$ and $\hat{\mathbf{R}}_k$ from the SDR solutions as
\begin{align}
\hat{\mathbf{w}}_{\mathrm{c}, k}=\frac{\tilde{\mathbf{R}}_k \mathbf{h}_k}{\sqrt{\mathbf{h}_k^{H} \tilde{\mathbf{R}}_k \mathbf{h}_k}},  \hat{\mathbf{R}}_k=\hat{\mathbf{w}}_{\mathrm{c}, k} \hat{\mathbf{w}}_{\mathbf{c}, k}^{H}, \quad  k=1, \cdots, K.
\label{cbf}
\end{align}
Given $\tilde{\mathbf{S}}_x$ and $\hat{\mathbf{R}}_k$, 
the sensing beamforming matrix $\hat{\mathbf{W}}_{\mathrm{s}}=\left[\hat{\mathbf{w}}_{\mathrm{s}, 1}, \hat{\mathbf{w}}_{\mathrm{s}, 2}, \cdots, \hat{\mathbf{w}}_{\mathrm{s}, N_t}\right]$ can be constructed from the SDR solutions using the Cholesky decomposition of the sensing Gram matrix as 
\begin{align}
\hat{\mathbf{W}}_{\mathrm{s}} \hat{\mathbf{W}}_{\mathrm{s}}^{H} = \hat{\mathbf{R}}_s = \tilde{\mathbf{S}}_x-\sum_{k=1}^K \hat{\mathbf{R}}_k .
\label{sbf}
\end{align}

We show in the following theorem that $\{ \tilde{\mathbf{S}}_x, \hat{\mathbf{R}}_k, k = 1,\cdots,K \}$ is a global optimum to problem (P2).

\color{blue}
\begin{theorem}
The solution $\{ \tilde{\mathbf{S}}_x, \hat{\mathbf{R}}_k, k = 1,\cdots,K \}$ is a global optimum to problem (P2).
\end{theorem}
\begin{proof}
See the appendix. 
\end{proof}
\color{black}

\section{Simulation Results and Analysis}

Suppose the largest possible target can be contained within a cubic region $D$ whose size is $1~\mathrm{m}  \times 1~\mathrm{m} \times 1~\mathrm{m}$. 
We choose the number of scatter points that constitute the target as $ N = 2048 $. 
Assume the BS is located at $(0,0,0)$~m and is responsible for sensing the target within the range of a 30~m radius sector on the horizontal plane, denoted by $ S = \{ (x , y , 0 ) \mid \operatorname{arctan}\frac{y}{x} \in [-60^\circ, 60^\circ] , \sqrt{x^2+y^2} \leq 30~\mathrm{m} \}$. 
The BS is equipped with a uniform linear array (ULA) with $ N_t = 64 $ transmitting antennas and a ULA with $ N_r = 64 $ receiving antennas. 
The transmitting and the receiving ULAs are both centered at $(0,0,0)$~m and are parallel to $y$ and $z$ directions, respectively, which is analogous to the Mills-Cross configuration \cite{Mills-cross}. 
The transmitting and the receiving antennas are set as dipoles polarized along $z$ and $y$ directions, respectively. 
\color{blue}
The carrier frequency is set as $f = 2.99$~GHz and the corresponding wavelength is $\lambda = 0.1 $~m. 
The inter-antenna spacing for both the transmitting and the receiving ULAs is set as $ \lambda/2 = 0.05 $~m. 
\color{black}
We assume the system operates with a bandwidth of $B = 10$~MHz,
and then all the thermal noise powers are set as $\sigma_s^2 = \sigma_k^2 = -174 + 10 \log _{10} B = -104 $ dBm, $k = 1,\cdots, K $. 
A total of $ L = 160 $ ISAC symbols are transmitted. 
Besides, the communications channel $\mathbf{h}_k$ is generated as i.i.d. CSCG variables $ \mathcal{C N}\left( \mathbf{0} , p_l \mathbf{I}_{N_t} \right) $ for all $k$.
The pathloss $ p_l $ is set as $ p_l = \frac{ G_t G_r \lambda^2 }{(4 \pi d)^2}$, where $G_t$ and $G_r$ are both set as $3$ dBi, 
and $d$ is uniformly and randomly sampled from $[50,150]$~m. 

For the diffusion model, we set the number of time steps as $ T = 200 $, and the diffusion coefficients $\beta_t$ linearly increase from $\beta_1 = 0.0001$ to $\beta_T = 0.05 $.
In the reverse noise estimator, we set $\bar{L} = 10$ and $L_{cs} = 10$.
The dimensions of the 10 concatsquash layers are set as 5-16-64-128-256-512-256-128-64-16-5, respectively.
\color{blue}
In order to train and test the diffusion model, we generate 100000 targets from the ShapeNet dataset \cite{shapenet}.
To handle the varying number of points in the ShapeNet dataset and ensure each target has exactly $N$ points, we use a fast uniform downsampling or upsampling strategy.
For point clouds with more than $N$ points, we uniformly downsample by removing points at regular intervals. For point clouds with fewer than $N$ points, we uniformly upsample by duplicating points at regular intervals.
\color{black}
The targets are uniformly and randomly located within the sector $S$. 
The reference location in the channel transferring network is set as $(x_r,y_r,z_r) = (3,0,0)$~m.  
The dataset is then split into training, testing, and validation sets by the ratio 80\%, 10\%, and 10\%, respectively.  
During the training process, we use the Adam optimizer and set the batch size as 256.   
Besides, we use the ground truth values of the sensing channel and the 
reference sensing channel for training.   
In order to compute the forward scattering, equation (\ref{lipp0}) is  
first converted into a discrete form by the methods of moments (MoM), and then the unknown total electric field $\mathbf{E}^t(\mathbf{r})$ is determined using the stabilized biconjugate gradient fast Fourier transform (BCGS-FFT) technique \cite{BCGS-FFT}.





Moreover, we introduce the mean Chamfer distance (MCD) between the ground truth and the estimated point clouds as the criterion to quantitatively evaluate the performance of EM property sensing, defined as
\begin{align} 
\mathrm{MCD} \! & = \! 10 \log_{10} \left[\ \frac{1}{| \mathcal{T}|} \sum_{\mathcal{X}^{(0)} \in \mathcal{T}} \left(  \frac{1}{N} \sum_{\mathbf{x} \in \mathcal{X}^{(0)}} \min _{\mathbf{y} \in \hat{\mathcal{X}}^{(0)}} \|\mathbf{x} - \mathbf{y}\|_2^2 \! \right. \right. \nonumber\\
& \left.\left. + \frac{1}{N} \sum_{\mathbf{y} \in \hat{\mathcal{X}}^{(0)}} \min _{\mathbf{x} \in \mathcal{X}^{(0)}}\|\mathbf{x}-\mathbf{y}\|_2^2 \right) \right],
\label{CD}%
\end{align} 
where $\mathcal{T}$ denotes the test dataset, and $ | \mathcal{T} | $ denotes the number of samples in the test dataset.




\begin{figure*}[t]
  \centering
\begin{minipage}[t]{0.33\linewidth}
\subfigure[Target real relative permittivity]{
\includegraphics[width=6cm,height=5.5cm]{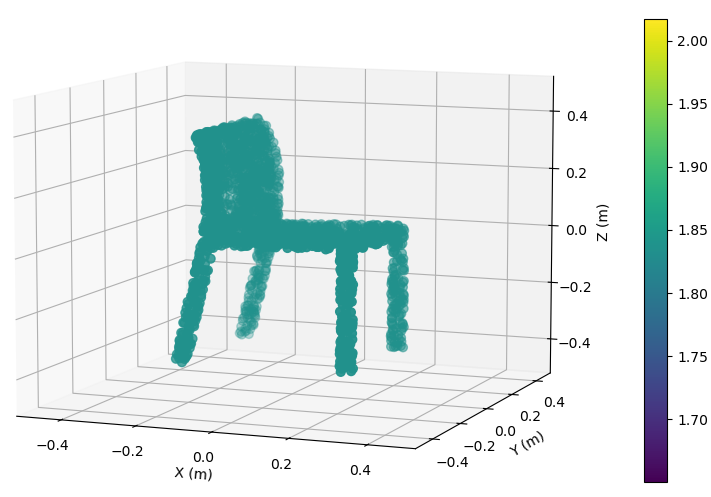}} 
\end{minipage}
\begin{minipage}[t]{0.33\linewidth}
\subfigure[Reconstructed relative permittivity \protect\\ with $P$ = 7 dBm]{
\includegraphics[width=6cm,height=5.5cm]{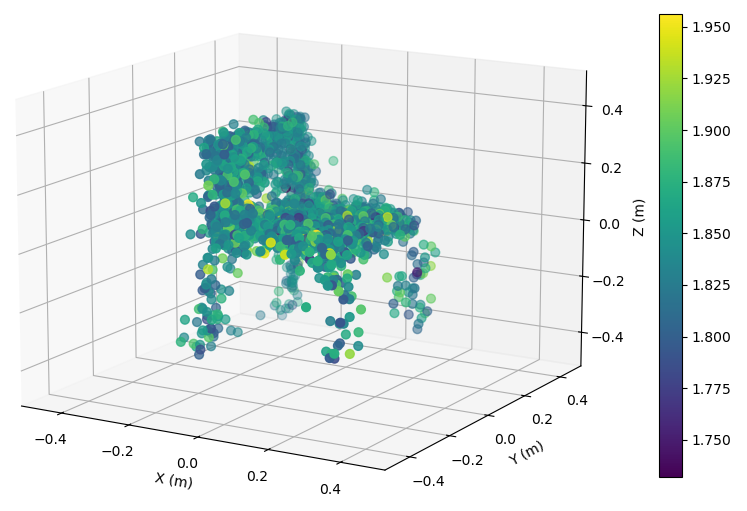}} 
\end{minipage} 
\begin{minipage}[t]{0.32\linewidth}
\subfigure[Reconstructed relative permittivity \protect\\ with $P$ = 15 dBm]{
\includegraphics[width=6cm,height=5.5cm]{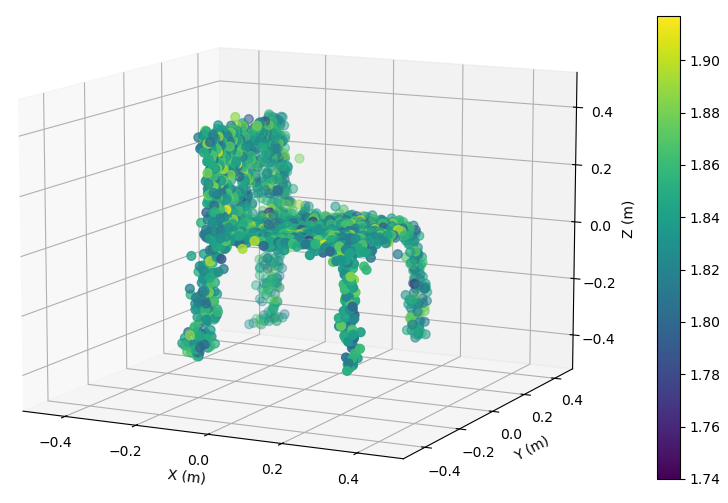}}   
\end{minipage} \\              
\begin{minipage}[t]{0.33\linewidth}
\subfigure[Target real conductivity]{
\includegraphics[width=6cm,height=5.5cm]{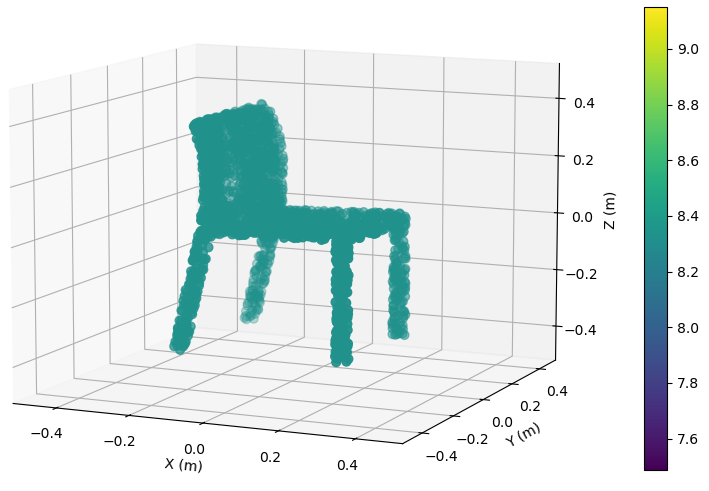}}
\end{minipage}
\begin{minipage}[t]{0.33\linewidth}
\subfigure[Reconstructed conductivity with $P$ = 7 dBm]{
\includegraphics[width=6cm,height=5.5cm]{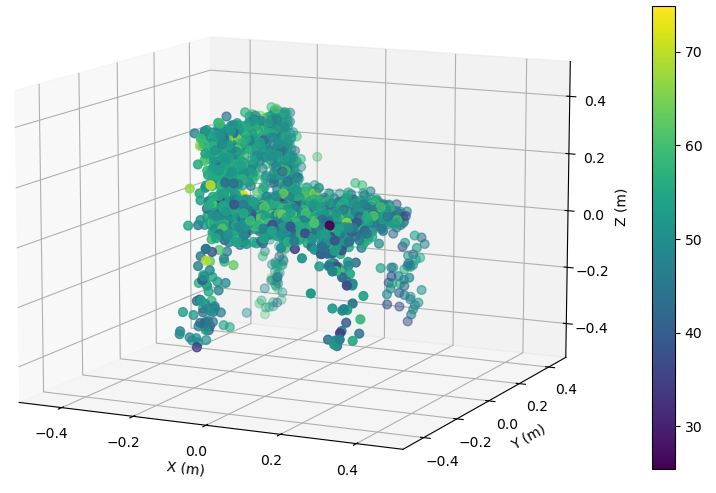}}
\end{minipage}
\begin{minipage}[t]{0.32\linewidth}
\subfigure[Reconstructed conductivity with $P$ = 15 dBm]{
\includegraphics[width=6cm,height=5.5cm]{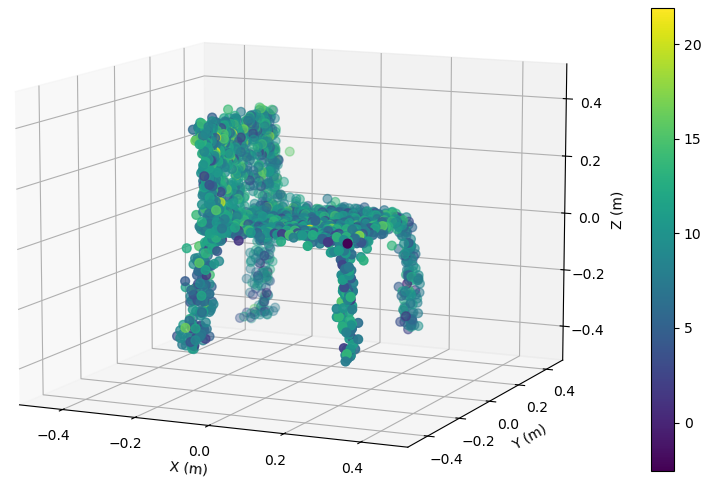}}
\end{minipage}
\caption{EM property sensing results versus $P$ with $K=2$ and $\bar{R}= 4$ bps/Hz. 
The center of the target is at $(15,0,0)$~m. 
The target is shown in the coordinate system relative to its center.
Unit of conductivity is mS/m. 
\textcolor{blue}{%
The target's real relative permittivity is 1.841, 
and the target's real conductivity is 8.325 mS/m.
The computational time to reconstruct the shown target is 0.2421 s and 0.2403 s with $P$ = 7 dBm and $P$ = 15 dBm, respectively.
}%
}
\label{image1}
\end{figure*}


\begin{figure}[t]
  \centering
\centerline{\includegraphics[width=8.4cm,height=6.5cm]{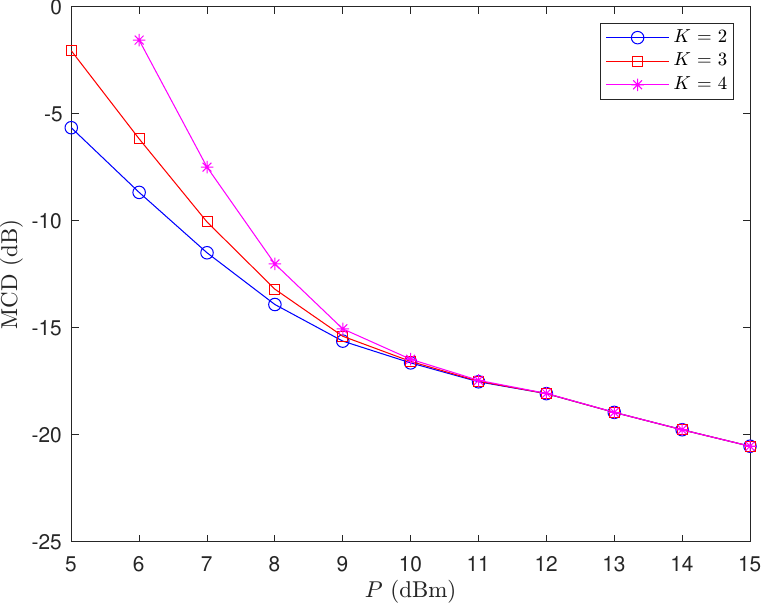}}
  \caption{MCD of 5D point clouds versus $P$ with $\bar{R} = 4$ bps/Hz. 
  The center of the target is at $(15,0,0)$~m.}
  \label{mcd_p}
\end{figure}


\subsection{EM Property Sensing Performance versus $P$}
To illustrate the EM property sensing results vividly, we present the reconstructed point clouds of the target based on relative permittivity and conductivity, respectively, with $K=2$ and $\bar{R}= 4$ bps/Hz in Fig.~\ref{image1}.
The center of the target is at $(15,0,0)$~m, and the target is shown in the coordinate system relative to its center.  
It is seen from Fig.~\ref{image1} that, the reconstructed point clouds can reflect the general shape of the target. 
The values of EM property reconstructed with $P = 15$ dBm is much more accurate compared to those reconstructed with $P = 7$ dBm.
Moreover, a higher $P$ value results in a more accurate reconstructed shape of the target.  
\color{blue}
The reason is that  a larger $P$ leads to a smaller error of the sensing channel estimation.
Since the estimated sensing channel is the latent of the diffusion model, 
a smaller error of $\hat{\mathbf{H}}_{s}$ further causes a smaller value of $\Delta \mathrm{VLB}$ and a 
smaller error of estimating \( \mathcal{X}^{(0)} \).
\color{black}

We explore the MCD of the reconstructed 5D point clouds versus $P$ in Fig.~\ref{mcd_p}. 
We set $\bar{R}= 4$ bps/Hz, and the center of the target is at $(15,0,0)$~m. 
It is seen from Fig.~\ref{mcd_p} that, the MCD decreases with the increase of $P$ in all cases, and the MCD is larger when $K$ is larger.
As more UEs need to be served, the power allocated to the sensing service is smaller,
which leads to a larger error of the sensing channel estimation and a larger MCD of the 5D point cloud reconstruction. 
The MCD decreases fast in the beginning when $P$ increases from $5$~dBm to $9$~dBm.  
When $P$ reaches a threshold of approximately $10$~dBm, the MCD reduces almost linearly in dB values with respect to $P$. 
When $P$ is relatively large, the discrepancy among different $K$ is not significant, which indicates that the transmit power can be mainly allocated to the EM property sensing service, and communications achievable rate limitations play a relatively small role in this period.


\begin{figure*}[t]
\captionsetup[subfigure]{singlelinecheck=false} 
\begin{minipage}[t]{0.33\linewidth}
\captionsetup[subfigure]{singlelinecheck=false}
\subfigure[Target relative permittivity]{
\includegraphics[width=6cm,height=5.5cm]{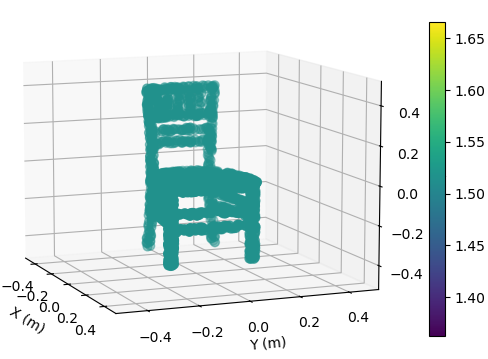}} 
\end{minipage}
\begin{minipage}[t]{0.33\linewidth}
\subfigure[Reconstructed relative permittivity with $\bar{R}$ = \\ 6 bps/Hz]{
\includegraphics[width=6cm,height=5.5cm]{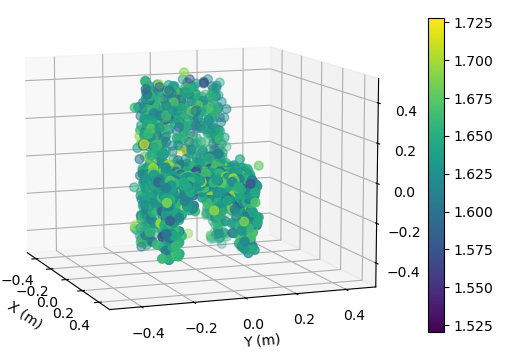}}  
\end{minipage}  
\begin{minipage}[t]{0.32\linewidth}
\subfigure[Reconstructed relative permittivity with $\bar{R}$ = \\ 4 bps/Hz]{
\includegraphics[width=6cm,height=5.5cm]{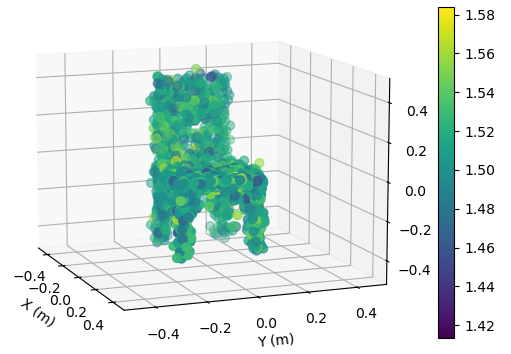}}   
\end{minipage} \\              
\begin{minipage}[t]{0.33\linewidth}
\captionsetup{singlelinecheck=false}
\subfigure[Target conductivity]{
\includegraphics[width=6cm,height=5.5cm]{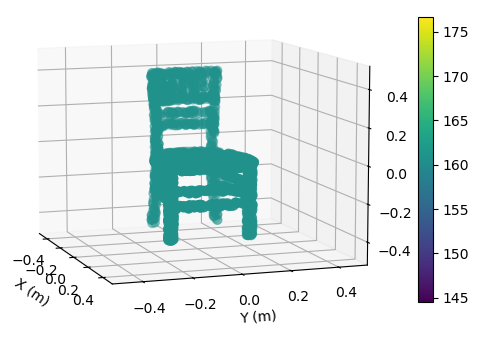}}
\end{minipage}
\begin{minipage}[t]{0.33\linewidth}
\subfigure[Reconstructed conductivity with $\bar{R}$ =  6 bps/Hz]
{\includegraphics[width=6cm,height=5.5cm]{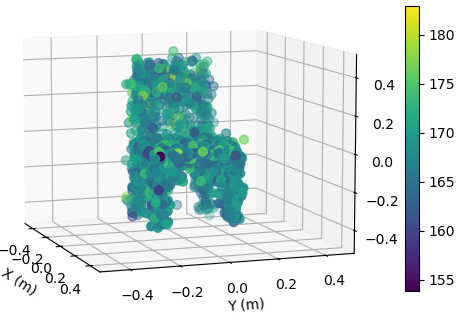}}
\end{minipage}
\begin{minipage}[t]{0.32\linewidth}
\subfigure[Reconstructed conductivity with $\bar{R}$ = 4 bps/Hz]{
\includegraphics[width=6cm,height=5.5cm]{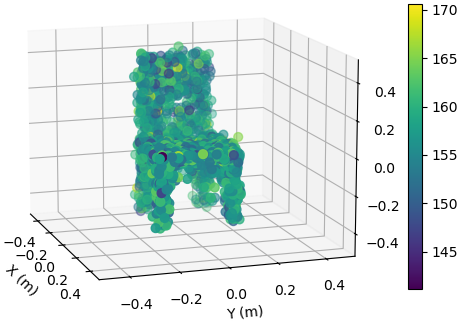}}
\end{minipage}
\caption{EM property sensing results versus $\bar{R}$ with $K = 2$ and $P=10$ dBm. 
The center of the target is at $(15,0,0)$~m. 
The target is shown in the coordinate system relative to its center.
Unit of conductivity is mS/m.
\textcolor{blue}{%
The target's real relative permittivity is 1.517, 
and the target's real conductivity is 160.8 mS/m.
The computational time to reconstruct the shown target is 0.2902 s and 0.2909 s with $\bar{R}$ = 6 bps/Hz and $\bar{R}$ = 4 bps/Hz, respectively.
}%
}
\label{image_K}
\end{figure*}


\begin{figure}[t]
  \centering
\centerline{\includegraphics[width=8.4cm,height=6.5cm]{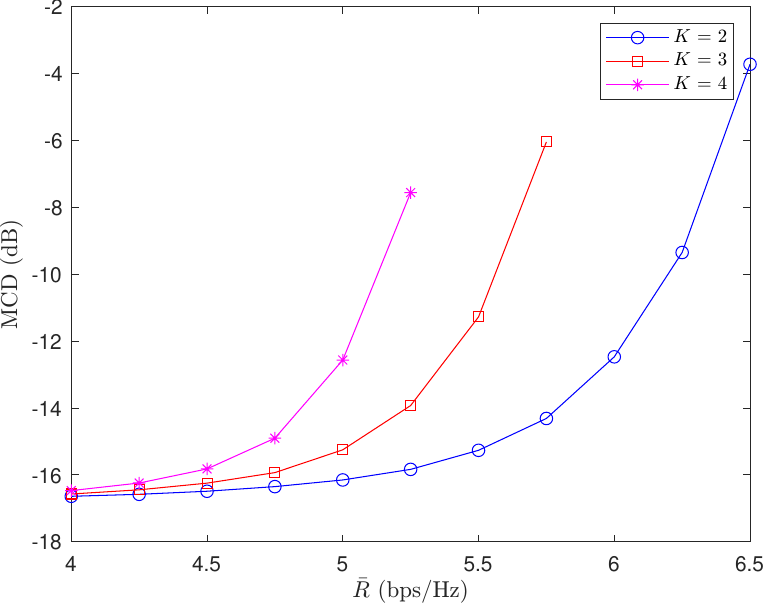}}
  \caption{\textcolor{black}{MCD of 5D point clouds versus $\bar{R}$ with $P = 10$ dBm. The center of the target is at $(15,0,0)$~m.}}
  \label{mcd_rate}
\end{figure}
\begin{figure*}[t]
\captionsetup[subfigure]{singlelinecheck=false} 
\begin{minipage}[t]{0.33\linewidth}
\captionsetup[subfigure]{singlelinecheck=false}
\subfigure[Target relative permittivity]{
\includegraphics[width=6cm,height=5.5cm]{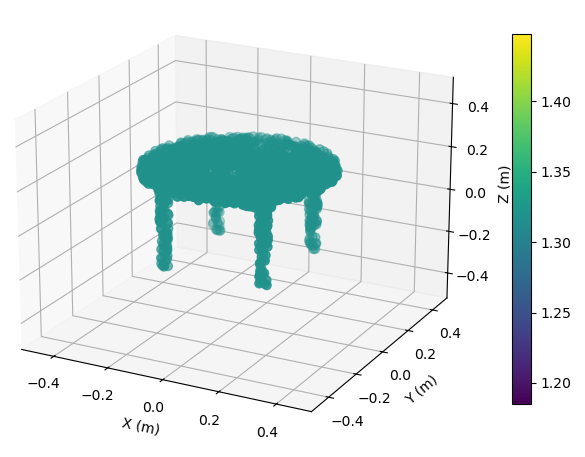}} 
\end{minipage}
\begin{minipage}[t]{0.33\linewidth}
\subfigure[Reconstructed relative permittivity with target at (25,0,0) m ]{
\includegraphics[width=6cm,height=5.5cm]{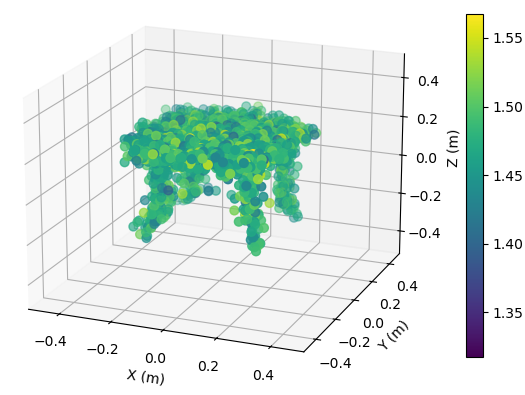}}  
\end{minipage}  
\begin{minipage}[t]{0.32\linewidth}
\subfigure[Reconstructed relative permittivity with target at (5,0,0) m ]{
\includegraphics[width=6cm,height=5.5cm]{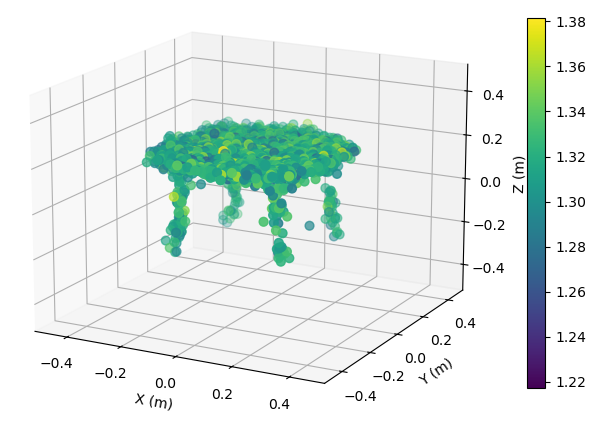}}   
\end{minipage} \\              
\begin{minipage}[t]{0.33\linewidth}
\captionsetup{singlelinecheck=false}
\subfigure[Target conductivity]{
\includegraphics[width=6cm,height=5.5cm]{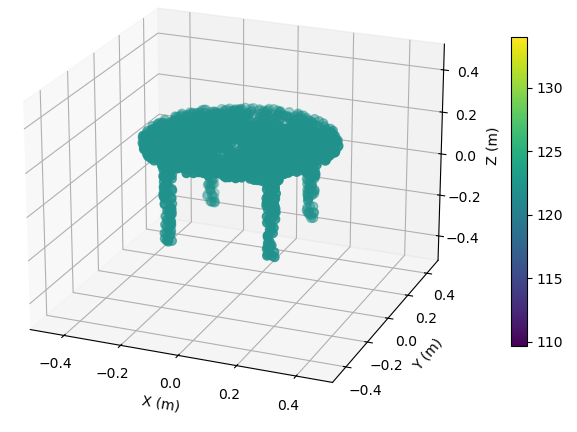}}
\end{minipage}
\begin{minipage}[t]{0.33\linewidth}
\subfigure[Reconstructed conductivity with target at (25,0,0) m ]
{\includegraphics[width=6cm,height=5.5cm]{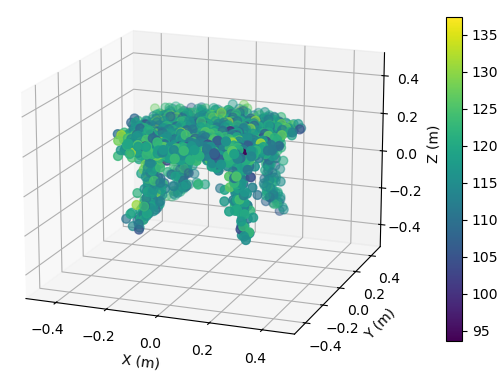}}
\end{minipage}
\begin{minipage}[t]{0.32\linewidth}
\subfigure[Reconstructed conductivity with target at (5,0,0) m ]{
\includegraphics[width=6cm,height=5.5cm]{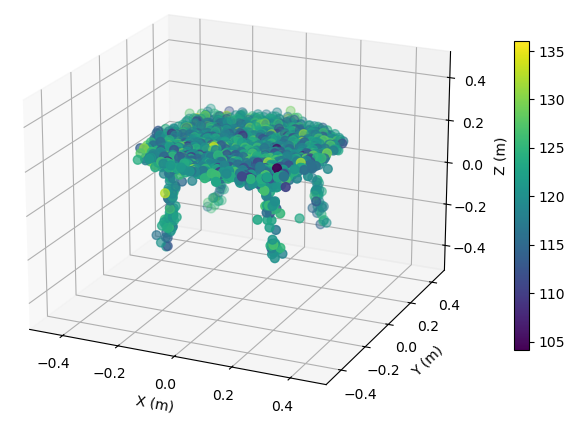}}
\end{minipage}
\caption{EM property sensing results versus location with $P=10$ dBm, $\bar{R} = 4$ bps/Hz, and $K = 2$. 
The target is shown in the coordinate system relative to its center.
Unit of conductivity is mS/m. 
\textcolor{blue}{%
The target's real relative permittivity is 1.318, 
and the target's real conductivity is 122.1 mS/m.
The computational time to reconstruct the shown target is 0.2331 s and 0.2416 s with target at (25,0,0) m and (5,0,0) m, respectively.
}%
}
\label{image_location}
\end{figure*}

\subsection{ EM Property Sensing Performance versus $\bar{R}$ }
To demonstrate the EM property sensing results, we present images of the target reconstructed based on relative permittivity and conductivity with $K=2$ and $P=10$ dBm. 
The center of the target is at $(15,0,0)$~m.  
As shown in Fig.~\ref{image_K}, the shape of the target reconstructed with $\bar{R} = 4$ bps/Hz is more accurate compared to that reconstructed with $\bar{R} = 6$ bps/Hz. 
Moreover, a smaller $\bar{R}$ yields more accurate reconstructed values of relative permittivity and conductivity. 
\color{blue}
The reason is that a smaller $\bar{R}$ leads to a larger power allocated to the sensing service and thus leads to 
a smaller error of the sensing channel estimation.
Since the estimated sensing channel is the latent of the diffusion model, 
a smaller error of $\hat{\mathbf{H}}_{s}$ further causes a smaller value of $\Delta \mathrm{VLB}$ and a 
smaller error of estimating \( \mathcal{X}^{(0)} \).
\color{black}

We investigate the MCD of the reconstructed 5D point clouds versus $\bar{R}$ with $P=10$ dBm in Fig.~\ref{mcd_rate}. 
The three lines are also known as the Pareto boundary of the MCD-rate region for different $k$. 
It is seen that, MCD increases with the increase of $\bar{R}$ in all cases, and the MCD is smaller with fewer communications UEs. 
When $\bar{R}$ is relatively low, the discrepancy among different $K$ is not pronounced, which indicates that the transmit power is mainly allocated to the EM property sensing service, and communications achievable rate limitations play a relatively small role in this period.
When $\bar{R}$ increases to a certain threshold, the minimum achievable rate can not be satisfied even when all transmit power is allocated to the communications service, which makes the feasible solution to problem (P1) nonexistent. 
When the number of UEs increases, the maximum feasible $\bar{R}$ decreases, because less power is allocated to the EM property sensing service. 

\subsection{ EM Property Sensing Performance versus Location of the Target}

\begin{figure}[t]
  \centering
\centerline{\includegraphics[width=8.4cm,height=6.5cm]{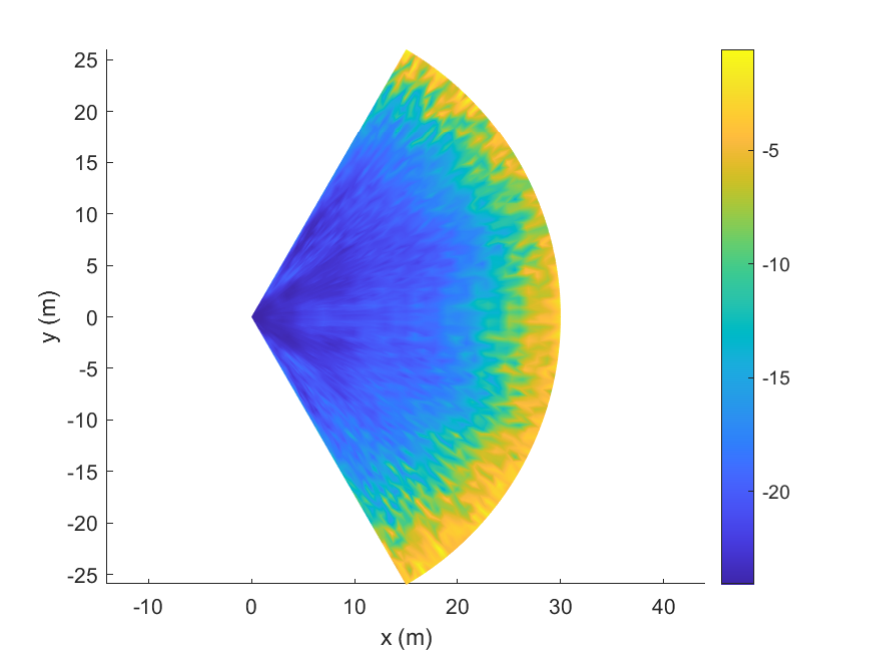}}
  \caption{\textcolor{black}{MCD of 5D point clouds versus location with $P=10$ dBm, $\bar{R} = 4$ bps/Hz, and $K = 2$.}}
  \label{mcd_location}
\end{figure}

To illustrate the EM property sensing results vividly, we present the reconstructed point clouds of the target based on relative permittivity and conductivity, respectively, with $P=10$ dBm, $\bar{R} = 4$ bps/Hz, and $K = 2$ in Fig.~\ref{image_location}.
The target is shown in the coordinate system relative to its center.  
It is seen from Fig.~\ref{image_location} that, the reconstructed 5D point clouds can reflect the general shape of the target. 
The values of EM property reconstructed with target at $(5,0,0)$~m is more accurate compared to those reconstructed with target at $(25,0,0)$~m. 
\color{blue}
Moreover, a closer distance results in a more accurate reconstructed shape of the target due to the following two reasons.
\begin{enumerate}
\item Firstly, when the target is closer to the BS, the sensing channel has more effective degrees of freedom (EDoF) \cite{myEDOF} and more diverse spatial features of the estimated sensing channel can be extracted by the channel transferring network in the diffusion model. 
\item Secondly, when the target is closer to the BS, the power of the echo signals becomes larger, which results in a larger sensing SNR. 
Thus, the NMSE of the reference sensing channel estimation is smaller, and the 5D point cloud reconstruction is more accurate accordingly. 
\end{enumerate}
\color{black}

We explore the MCD of the reconstructed 5D point clouds versus the location of the target in Fig.~\ref{mcd_location}. 
We set $P=10$ dBm, $\bar{R} = 4$ bps/Hz, and $K = 2$. 
It is seen from Fig.~\ref{mcd_location} that, the MCD is generally smaller when the target is in closer proximity to the BS. 
The MCD does not vary significantly with the change of the angle, which indicates that the proposed method can sense the EM property of the target effectively in any direction of the sector $S$.

\color{blue}
\subsection{ Generalization to Different Scenarios }
\begin{figure}[t]
  \centering 
\centerline{\includegraphics[width=8.4cm,height=6.5cm]{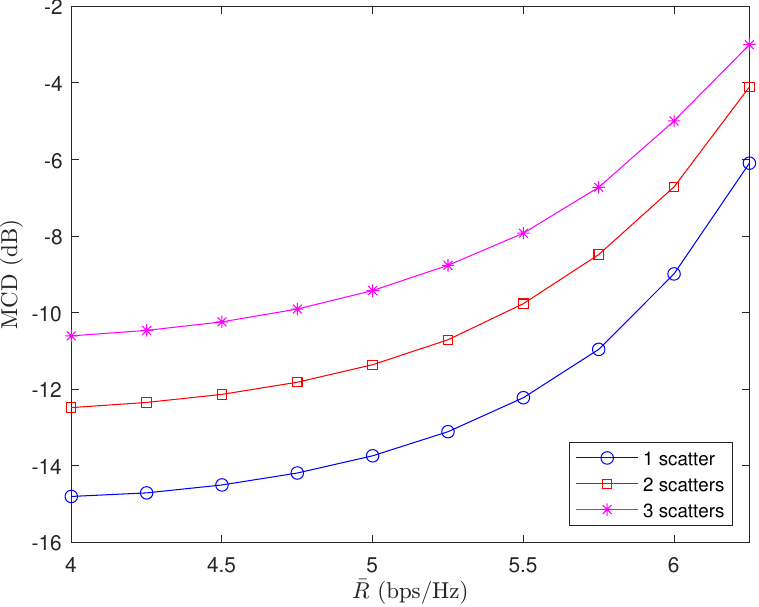}}
  \caption{\textcolor{blue}{MCD of 5D point clouds in different scenarios with $P=10$ dBm, $\bar{R} = 4$ bps/Hz, $K = 2$, and different numbers of scatterers close to the target. The center of the target is at $(15,0,0)$~m.}}
  \label{mcd_sce}
\end{figure}

In order to explore the proposed model's generalizability to different scenarios, we now test the model in 3 different scenarios with 3 different numbers of  scatterers close to the target. The center of the target is at $(15,0,0)$~m, and the scatterers are sequentially located at $(15,5,0)$~m, $(12.5,-4.33,0)$~m, and $(17.5,-4.33,0)$~m, respectively. The radar cross section of each scatterer is set as $0.1$ $\mathrm{m}^2$.
We explore the MCD of the reconstructed 5D point clouds versus $\bar{R}$ with $P=10$ dBm, $\bar{R} = 4$ bps/Hz, $K = 2$, and different numbers of  scatterers close to the target in Fig.~\ref{mcd_sce}. 
It is seen from Fig.~\ref{mcd_sce} that, the MCD increases with the increase of $\bar{R}$ in all cases, and the MCD is smaller with fewer nearby scatterers. 
The reason is that when the number of scatterers increases, the power of the interference signals echoed back from the scatterers becomes larger, and the MCD of 5D point cloud reconstruction consequently becomes larger. 

\subsection{ Generalization to Different Carrier Frequency }
\begin{figure}[t]
  \centering
\centerline{\includegraphics[width=8.4cm,height=6.5cm]{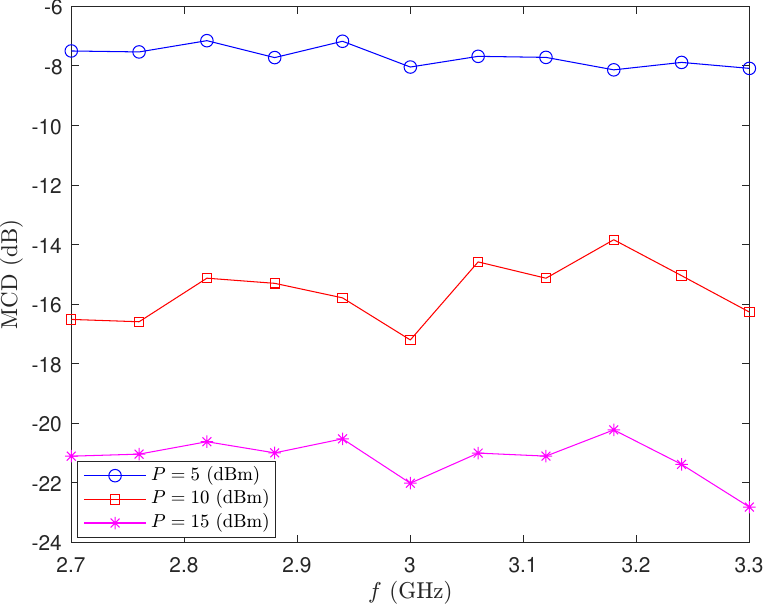}}
  \caption{\textcolor{blue}{MCD of 5D point clouds versus carrier frequency $f$ with $\bar{R} = 4$ bps/Hz and $K = 2$. The center of the target is at $(15,0,0)$~m.}}
  \label{mcd_freq}
\end{figure}

In order to explore the proposed model's generalizability to different carrier frequency, we expand the dataset to 11 uniformly-spaced carrier frequencies between 2.7 GHz and 3.3 GHz, where the process to generate the dataset of each frequency is the same as that described in the second paragraph of Section~\uppercase\expandafter{\romannumeral6}. 
We train the model on 2.7 GHz, 3 GHz, and 3.3 GHz and test the model on all 11 carrier frequencies. 
We explore the MCD of the reconstructed 5D point clouds versus carrier frequency $f$ with $\bar{R} = 4$ bps/Hz and $K = 2$ in Fig.~\ref{mcd_freq}. 
The center of the target is at $(15,0,0)$~m, and there is no other scatterer. 
It is seen from Fig.~\ref{mcd_freq} that, the MCD decreases with the increase of $P$ in all cases.
The MCD does not vary significantly with the change of $f$, which indicates that the proposed method can sense the EM property of the target effectively when the system operates at different carrier frequencies.
\color{black}

\section{Conclusion}
In this paper, we propose a groundbreaking ISAC scheme that harnesses the power of the diffusion model to precisely capture the EM property of a target within a predetermined sensing area. The innovative approach begins with the estimation of the sensing channel using both communications and sensing signals. Subsequently, the diffusion model is applied to create a comprehensive point cloud, offering a vivid 3D depiction of the target's EM property distribution, encompassing essential attributes such as shape, relative permittivity, and conductivity. 
We further emphasize the importance of high-fidelity reconstruction by designing communications and sensing beamforming matrices that aim to minimize the MCD between the actual and the estimated point clouds. This optimization is meticulously conducted under the constraints of maximum transmit power and a minimum communications rate for UE.  
The simulation results demonstrate the effectiveness of the proposed method to realize high-quality reconstruction of the target's EM property.
Moreover, the method's robustness in sensing the EM property of the target, regardless of its position within the sensing area, demonstrates its versatility and potential for various practical applications requiring accurate EM property sensing.
The proposed ISAC scheme stands as a significant advancement in the field of wireless communications and sensing, paving the way for more precise and reliable EM property assessments.
\color{blue}
\appendix

\section{Proof of Theorem 1}
\label{app:proofs} 
We first prove that $\tilde{\mathbf{S}}_x$ and $\hat{\mathbf{R}}_k$ satisfy constraints (\ref{36b})-(\ref{36f}) to ensure they are feasible solutions to problem (P2).
It can be easily seen that constraints (\ref{36c})-(\ref{rank1}) hold obviously with (\ref{cbf}) and (\ref{sbf}). 
Therefore, we only need to check the remaining constraints (\ref{36b}) and (\ref{36f}). 

Constraint (\ref{36b}) holds with $\hat{\mathbf{R}}_k$, because there is $\mathbf{h}_k^{H} \hat{\mathbf{R}}_k \mathbf{h}_k=\mathbf{h}_k^{H} \tilde{\mathbf{R}}_k \mathbf{h}_k$ and (\ref{36b}) holds with $\tilde{\mathbf{R}}_k$.
In order to check constraint (\ref{36f}), we only need to prove that $\tilde{\mathbf{R}}_k-\hat{\mathbf{R}}_k \succeq 0$. 
Based on the Cauchy-Schwarz inequality, for any $\mathbf{u} \in \mathbb{C}^{ N_t \times 1 }$, there is 
\begin{align}
\left(\mathbf{h}_k^{H} \tilde{\mathbf{R}}_k \mathbf{h}_k\right) \!\!  \left(\mathbf{u}^{H} \tilde{\mathbf{R}}_k \mathbf{u}\right) \!= \!\left\|\tilde{\mathbf{R}}_k^{\frac{1}{2}} \mathbf{h}_k \right\|_2^2 \left\|\tilde{\mathbf{R}}_k^{\frac{1}{2}} \mathbf{u}  \right\|_2^2 
\geq 
\!\left|\mathbf{u}^{H} \tilde{\mathbf{R}}_k \mathbf{h}_k\right|^2\! \! , 
\label{cauchy}
\end{align}
where $\tilde{\mathbf{R}}_k^{\frac{1}{2}}$ is the Cholesky decomposition of $ \tilde{\mathbf{R}}_k $ that satisfies $(\tilde{\mathbf{R}}_k^{\frac{1}{2}})^{H} \tilde{\mathbf{R}}_k^{\frac{1}{2}} = \tilde{\mathbf{R}}_k$. 
Substituting (\ref{cbf}) into $\hat{\mathbf{R}}_k$ and then utilizing (\ref{cauchy}), we have 
\begin{align}
\mathbf{u}^{H}\left(\tilde{\mathbf{R}}_k-\hat{\mathbf{R}}_k\right) \mathbf{u} & =\mathbf{u}^{H} \tilde{\mathbf{R}}_k \mathbf{u}  -\frac{\left|\mathbf{u}^{H} \tilde{\mathbf{R}}_k \mathbf{h}_k\right|^2}{\mathbf{h}_k^{H} \tilde{\mathbf{R}}_k \mathbf{h}_k} \geq 0 . 
\label{arbitra}
\end{align}

Due to the arbitrariness of $\mathbf{u} \in \mathbb{C}^{ N_t \times 1 }$  in (\ref{arbitra}), there is $\tilde{\mathbf{R}}_k-\hat{\mathbf{R}}_k \succeq 0$. 
Since there is $\tilde{\mathbf{S}}_x-\sum_{k=1}^K \tilde{\mathbf{R}}_k \succeq 0$, we have 
\begin{align}
\tilde{\mathbf{S}}_x-\sum_{k=1}^K \hat{\mathbf{R}}_k=\tilde{\mathbf{S}}_x-\sum_{k=1}^K \tilde{\mathbf{R}}_k+\sum_{k=1}^K\left(\tilde{\mathbf{R}}_k-\hat{\mathbf{R}}_k\right) \succeq 0  . 
\end{align}
Therefore, we have proven that $\{ \tilde{\mathbf{S}}_x, \hat{\mathbf{R}}_k, k = 1,\cdots,K \}$ is a feasible solution to problem (P2). 

Since $\{ \tilde{\mathbf{S}}_x, \tilde{\mathbf{R}}_k, k = 1,\cdots,K \}$ is a global optimum to problem (P2) without the rank-1 constraint and the objective function has the same value for $\{ \tilde{\mathbf{S}}_x, \tilde{\mathbf{R}}_k, k = 1,\cdots,K \}$ and $\{ \tilde{\mathbf{S}}_x, \hat{\mathbf{R}}_k, k = 1,\cdots,K \}$, $\{ \tilde{\mathbf{S}}_x, \hat{\mathbf{R}}_k, k = 1,\cdots,K \}$ is a global optimum to problem (P2), which completes the proof. 
\color{black}

 \small 
 \bibliographystyle{ieeetr}
 \bibliography{IEEEabrv,mainbib}

\ifCLASSOPTIONcaptionsoff
  \newpage
\fi



%

\end{document}